\documentclass[aps,prd,onecolumn,notitlepage,showpacs,showkeys,superscriptaddress,nofootinbib,11pt]{revtex4-2}

\def \be {\begin{equation}} 
\def \ee {\end{equation}} 
\def \bea {\begin{eqnarray}} 
\def \eea {\end{eqnarray}} 

\usepackage{physics}
\usepackage{commath}
\usepackage{url}
\usepackage{bm}
\usepackage{braket}
\usepackage{accents}

\usepackage{graphicx}
\usepackage{dcolumn}
\usepackage{bm}
\usepackage{epsfig} 
\usepackage{amsfonts}
\usepackage{amsmath}
\usepackage{amssymb}
\usepackage[usenames]{color}
\usepackage[dvipsnames]{xcolor}
\usepackage[unicode, colorlinks=true, linkcolor=linkcolor, citecolor=linkcolor, filecolor=linkcolor,urlcolor=linkcolor, pdfusetitle]{hyperref}
\usepackage{listings}
\usepackage[T1]{fontenc}
\usepackage{setspace}

\hypersetup{colorlinks,citecolor=blue,linkcolor=blue,urlcolor=blue}
\hypersetup{final=true}
\graphicspath{figs}

\begin{document}
\SetSinglespace{1.25}
\singlespacing

\title{Examining the validity of the minimal varying speed of light model through cosmological observations: relaxing the null curvature constraint}

\author{Purba Mukherjee}
\email{purba16@gmail.com}
\affiliation{Physics and Applied Mathematics Unit, Indian Statistical Institute, Kolkata - 700108, India}

\author{Gabriel Rodrigues}
\email{gabrielrodrigues@on.br}
\affiliation{Observat\'orio Nacional, 20921-400, Rio de Janeiro - RJ, Brazil}

\author{Carlos Bengaly}
\email{carlosbengaly@on.br}
\affiliation{Observat\'orio Nacional, 20921-400, Rio de Janeiro - RJ, Brazil}

\begin{abstract}

We revisit a consistency test for the speed of light variability, using the latest cosmological observations. This exercise can serve as a new diagnostics for the standard cosmological model and distinguish between the minimal varying speed of light in the Friedmann-Lema\^{i}tre-Robertson-Walker universe. We deploy Gaussian processes to reconstruct cosmic distances and ages in the redshift range $0<z<2$ utilizing the Pantheon compilation of type-Ia supernova luminosity distances (SN), cosmic chronometers from differential galaxy ages (CC), and measurements of both radial and transverse modes of baryon acoustic oscillations ($r$-BAO and $a$-BAO) respectively. Such a test has the advantage of being independent of any non-zero cosmic curvature assumption  - which can be degenerated with some variable speed of light models - as well as any dark energy model. We also examine the impact of cosmological priors on our analysis, such as the Hubble constant, supernova absolute magnitude, and the sound horizon scale. We find null evidence for the speed of light variability hypothesis for most choices of priors and data-set combinations. However, mild deviations are seen at $\sim 2\sigma$ confidence level for redshifts $z<1$ with some specific prior choices when $r$-BAO data is employed, and at $z>1$ with a particular reconstruction kernel when $a$-BAO data are included. Still, we ascribe no statistical significance to this result bearing in mind the degeneracy between the associated priors for combined analysis, and incompleteness of the $a$-BAO data set at higher $z$. 

\end{abstract}

\vskip 1.0cm



\maketitle

\vskip 1.0cm

%%%%%%%%%%%%%%%%%%%%%%%%%%%%%%%%%%%%%%%%%%%%%%%%%%%

\section{Introduction}

The standard cosmological model (SCM), corresponds to the flat-$\Lambda$CDM paradigm. Based on the assumption of a spatially flat, homogeneous and isotropic universe with Einstein's general relativity as the correct description of gravity, such a scenario can accurately explain most of the observational data at astrophysical and cosmological scales \cite{peebles,copeland}. Since the discovery of late-time accelerated expansion of the Universe in the late 1990s~\cite{riess98, perlmutter99}, it provides the best explanation to the cosmological observations of the Cosmic Microwave Background (CMB)~\cite{planck21}, type Ia Supernovae (SN-Ia) luminosity distances~\cite{pantheon18}, and the clustering and weak lensing of cosmic structures~\cite{eboss21, kids21, des21a, des21b}, even today. 
		
Despite being so well-established today, the $\Lambda$CDM model is plagued with some unresolved issues, i.e., the value of the vacuum energy density~\cite{weinberg89,padmanabhan03}, the primordial singularity problem~\cite{baumann18}, and observational discrepancies between measurement of cosmological parameters, arising from different scales of observations~\cite{H0tension,s8tension}. One of the most prominent and intriguing problems, at this moment, is the $\sim 5\sigma$ tension~\cite{divalentino21, riess22, shah21} between the present value of Hubble parameter $H_0$ measured in the late-time by the SH0ES \cite{riess22} team, and the inferred value from the early CMB sky by the Planck \cite{planck21} survey assuming a $\Lambda$CDM background. Therefore, it is crucial to revisit the foundations of SCM and look for alternative solutions or mechanisms to overcome these problems. 
			
The existence of fundamental constants is a keystone governing the laws of physics. So, as an unconventional route one can probe the constancy of these physical constants, based on the assumption that they may have varied at some point in the past or still are varying in the present \cite{uzan}. For instance, the constancy of the speed of light $c$, which not only shapes the fundamental basis of Einstein's general relativity (as well as other theories of gravity) but also plays a fundamental role in Electromagnetism and Thermodynamics, thus being one of the most important and fundamental postulates in physics. Therefore, an assessment of the constancy of $c$ across different epochs of the Universe is a strong test of fundamental physics, and any measurement other than a constant would require a profound reformulation of the physics that underlies this fundamental postulate. 

The impact of a variable speed of light (VSL) on Physics and Cosmology has been discussed by~\cite{dicke57, moffat93, magueijo00, avelino99, moffat16}. In recent years, VSL theories have been developed for the sake of solving some theoretical caveats at primordial times, e.g. the horizon and flatness problems, in addition to the cosmic coincidence and the $H_0$ tension in the late-time Universe~\cite{moffat93, barrow98, albrecht99, barrow99a, barrow99b, clayton99, avelino99, clayton00, bassett00, magueijo00, clayton02, magueijo03, ellis05, ellis07, magueijo08, cruz12, moffat16, franzmann17, cruz18, costa19, gupta20, lee21a, lee21b, lee21c, lee21d, lee21e, cuzinatto22, lee23, gupta23}. Some of these theories may provide viable solutions to these SCM issues, albeit must reproduce the success of special relativity in explaining electromagnetism and thermodynamics, at least.  

Although the speed of light has been measured with extremely high precision and no significant variation has yet been found, an absolute majority of them were carried out on the Earth and Solar System scales. We still do not have many measurements arising from the distant Universe, given the difficulty in obtaining and extracting cosmological data in higher redshifts. However, in recent years, a significant improvement in both the quality and quantity of observations at cosmological scales has allowed some progress in this field. 

A method to measure a possible variation of the speed of light through Baryon Acoustic Oscillations (BAO) was proposed by Salzano \textit{et al}~\cite{salzano15}. This method relies on a relation between the maximum value of the angular diameter distance $d_A(z)$ and corresponding value of the Hubble parameter $H(z)$ at some redshift $z_m$ where this maximum occurs, $c(z_m)=d_A(z_m)H(z_m)$. Cao \textit{et al}~\cite{cao17} carried out an estimation for the speed of light $c = (3.039\pm 0.180)\;\times 10^5\; \mathrm{km\;s}^{-1}$ using the reconstructed angular diameter distances through Gaussian processes from intermediate-luminosity radio quasars, calibrated as standard rulers, and the Hubble parameter from cosmic chronometers at the maximum redshift of $z_m=1.7$. 

Recently, Rodrigues \& Bengaly~\cite{gabriel22} carried out an estimation for the speed of light, using the Pantheon SN-Ia compilation to reconstruct the angular diameter distance through the cosmic distance duality relation, combined with the Hubble parameter observations from cosmic chronometers (CC) and the radial mode of baryonic acoustic oscillations (BAO) to reconstruct the Hubble diagram. These reconstructions were undertaken though Gaussian processes, reporting two $\sim 5$\% precision measurements of $c = (3.20 \pm 0.16) \; \times 10^5 \; \mathrm{km \; s}^{-1}$ in the maximum redshift $z_m \simeq 1.58$, and $c = (2.67 \pm 0.14) \; \times 10^5 \; \mathrm{km \; s}^{-1}$ in the maximum redshift $z_m \simeq 1.36$, depending on the kernel used. This prescription has its own limitations, as we still do not have ample data available around the redshift where the angular diameter distance typically reaches the maximum value, thus providing results with limited precision. Moreover, this test assumes a null curvature value, which can be degenerated with some VSL models, as explored in Salzano~\cite{salzano17}, where the author developed a method to measure $c$ at multiple redshifts.  

In Cai {\it et al}~\cite{cai16}, a consistency test of the speed of light variability was initially proposed reporting null evidence for VSL in their analysis. In this work, we revisit this test, independent of any non-zero cosmic curvature assumption - which can be degenerated with VSL models, in light of the latest cosmological observations. This provides a more robust method to probe fundamental physics breakdown at such cosmic scales. We further investigate the role played by the different cosmological parameters involved in the analysis in view of the growing tensions between diverse observational probes.
		
This paper is organized as follows. In section 2, we describe the theoretical framework of our analysis. Section 3 presents the methodology adopted and briefly reviews the observational data. The results obtained are presented in section 4. Finally, we conclude the manuscript in Section 5 with an overall discussion about the results.

%%%%%%%%%%%%%%%%%%%%%%%%%%%%%%%%%%%%%%%%%%%%%%%%%%%

\section{Theoretical Framework}

The Universe on a large scale is described by the spatially homogeneous and isotropic Friedmann-Lema\^{i}tre-Robertson-Walker (FLRW) metric, 
\begin{equation}\label{eq:flrw}
\small \dif s^2 = - c^2 \dif t^2 + a^2(t) \left[\frac{\dif r^2}{1-kr^2} + r^2 \dif \theta^2 + r^2 \sin^2\theta \dif \phi^2 \right], 
\end{equation}
where $a(t)$ is the scale factor and $k$ is the curvature index. The isotropy and homogeneity of the 3D-space section demand $k$ to be a constant, which can thus be scaled to pick up values from $+1, -1, 0$ corresponding to closed, open or flat space sections, respectively.  

\noindent In FLRW Universe, assuming that the speed of light is time-dependent, i.e., $c \equiv c(t)$,  one can arrive at the expression for the proper distance starting from Eq. \eqref{eq:flrw}, as
\begin{align} 
&\frac{\dif r}{\sqrt{1-kr^2}} = \pm \frac{c \dif t}{a} \\
\implies \int_{0}^{r} &\frac{\dif r^\prime}{\sqrt{1-k{r^\prime}^2}} = \int_{0}^{t_0} \frac{c(t^\prime) \dif t^\prime }{a(t^\prime)} = \int_{a}^{1} \frac{c(a^\prime) \dif a^\prime}{{a^\prime}^2 H(a^\prime)} = \int_{0}^{z} \frac{c(z^\prime) \dif z^\prime}{H(z^\prime)},
\end{align}
where $H = \frac{\dot {a}}{a}$ is the Hubble parameter. 

\noindent Now, if \begin{equation} \label{eq:c}
c(z) = c_0 \hat{c}(z) ,
\end{equation}
is the time-varying speed of light, then 
\begin{equation} \label{eq:int_r}
\int_{0}^{r} \frac{\dif r^\prime}{\sqrt{1-k{r^\prime}^2}} = \frac{c_0}{H_0} \int_{0}^{z} \frac{\hat{c}(z^\prime) \dif z^\prime}{E(z^\prime)},  
\end{equation}
where
\begin{equation}\label{eq:dp}
D_p (z)= \int_{0}^{z} \frac{\hat{c}(\tilde{z})}{E(\tilde{z})} \dif \tilde{z} ,
\end{equation} 
is the normalized proper distance from the observer to a celestial object at redshift $z$ along the line of sight, and \begin{equation}\label{eq:Hubble_reduced}
E(z) = \frac{H (z)}{H_0},
\end{equation} 
is the reduced Hubble parameter. It should be noted that any suffix 0 indicates values of the relevant quantities at the present epoch and $z$ is the redshift, defined as $1+z \equiv a/a_0$. Throughout our work, we consider the speed of light at $z=0$ as $c_0 = 3\times 10^5$ km s$^{-1}$. 

\noindent Note that these relations are only valid for some classes of time-varying speed of light models, as for instance, the VSL model solutions presented in~\cite{barrow98}, often referred to as the minimal varying speed of light (mVSL) models. Other cases, e.g. the so-called minimally extended varying speed of light (meVSL) models~\cite{lee21a}, present different FLRW solutions, so they will not be considered in our analysis.

\noindent The transverse comoving distance $d_C(z)$ is obtained from integrating the left-hand side of Eq. \eqref{eq:int_r}, 
\begin{equation}
	\frac{1}{\sqrt{\vert k \vert}} S_k^{-1} \left[ \sqrt{\vert k \vert} ~r(z) \right] = \frac{c_0}{H_0} D_p(z) ~, ~~~ \sqrt{\vert k \vert} \equiv \frac{ \sqrt{\vert \Omega_{k0} \vert} H_0}{c_0}.
\end{equation} 
On rearranging, we arrive at
\begin{equation} \label{eq:dc_vs_D}
	r(z) \equiv d_C(z) = \frac{c_0}{H_0} D(z) ,
\end{equation}
where $D(z)$ is the normalized transverse comoving distance, given by
\begin{equation} \label{eq:D}
D(z)= \frac{1}{\sqrt{\vert \Omega_{k0} \vert}} S_k \left[ \sqrt{\vert \Omega_{k0} \vert} \int_{0}^{z} \frac{\hat{c}(z^\prime) \dif z^\prime}{E(z^\prime)}\right] ,
\end{equation}
in which the $S_k$ function is a shorthand for,
\[ S_k \left[ x \right] = \begin{cases}
\sinh x &  (k = -1), \\
~~x & (k = ~0), \\
\sin x & (k = +1).
\end{cases}
\]

Differentiating Eq. \eqref{eq:D} with the redshift $z$,  we get \begin{equation}
\Omega_{k0} = \frac{{D^\prime}^2 E^2 - \hat{c}^2}{D^2 \hat{c}^2} ~~~ \text{for both  }~ \Omega_{k0} \neq 0.
\label{equa:OmegaK}
\end{equation} 

For eliminating $\Omega _{k0}$, we utilize the second derivative of $D(z)$ and arrive at,
	\begin{equation} \label{eq:null-test}
	{\hat c^3}(z) + A(z){\hat c}(z) + B(z)\hat c'(z) = 0
	\end{equation}
	where
	\begin{equation}\label{eq:Az}
	A(z) = \left[ D''(z)E(z) + D'(z)E'(z) \right] D(z)E(z) - D{'^2}(z){E^2}(z),
	\end{equation}
	and 
	\begin{equation}\label{eq:Bz}
	B(z) = - D(z)D'(z){E^2}(z) .
	\end{equation}

Note that, the Eq. \eqref{equa:OmegaK} holds only for $\Omega_{k0} \neq 0$. Thus, the above
constancy condition given by Eq. \eqref{eq:null-test} is valid only for non-flat universes. 

On assuming the constancy of the speed of light, i.e., $\hat c(z)=c(z)/c_0=1$, Eq. \eqref{eq:null-test} reduces to,
\begin{eqnarray} \label{eq:Tz}
T(z) &\equiv& 1 + A(z) = 0 \,; \\ \notag
T(z) &\neq& 0 \;\; \text{implies SCM ruled out.}
\end{eqnarray}
Any deviation of $T$ from $0$ at some redshift $z_*$ indicates $c({z_*})$ is different from $c_0$. Hence, Eq. \eqref{eq:Tz} serves as the key equation for our analysis.

\section{Methodology}

\subsection{Observational Datasets}

In this work, we use different combinations of datasets like the Cosmic Chronometer~\cite{cc0,cc1,cc2,cc3,cc4,cc5,cc6,cc7,cc8,cc9} Hubble data, the Pantheon~\cite{pantheon18} compilation of the Type Ia Supernova apparent magnitudes, the radial mode of Baryon Acoustic Oscillation~\cite{bao1,bao2,bao3,bao4,bao5,bao6,bao7,bao8,bao9} (hereafter $r$-BAO) measurements, and the transverse angular-BAO distance measurements (hereafter referred to as $a$-BAO). The latter consists on 15 transverse BAO mode ($\theta_{\text{BAO}}$) data points taken from Table I of~\cite{nunes20}, as obtained from SDSS-III luminous red galaxies~\cite{carvalho16, alcaniz17, carvalho20} and quasars~\cite{decarvalho18}, besides SDSS-IV blue galaxies~\cite{decarvalho21}. So we can reconstruct $T(z)$, defined in Eq.\eqref{eq:Tz}, as a function of the redshift $z$. 

The Cosmic Chronometer $H(z)$ measurements depend on the differential ages between galaxies and do not assume any particular cosmological model. 

The SNe-Ia data apparent magnitude does not depend on cosmological assumptions as well. However, their absolute magnitude does rely on the Chandrasekhar mass limit, which depends on fundamental constants such as $c_0$, Newton's gravitational constant $G$ and Planck's constant $h$. We will not revisit how this quantity could be affected by varying fundamental constants in this work, as our plan is to assume these absolute magnitude measurements as priors in our null test. We refer an interested reader to~\cite{gupta23}  for further details on how fundamental astrophysical relations (such as the Chandrasekhar mass limit itself) could be modified in a model that allows covariation of fundamental constants. Also note that a late-time transition in $G$ was recently proposed in the literature as a possible solution for the $H_0$ tension~\cite{kazantzidis20, ruchika23}. 

The radial BAO peaks in the galaxy power spectrum, or the Ly-$\alpha$ forest of QSOs, give an alternative compilation $H(z)$ based on the clustering. Now, there are apprehensions that the $r$-BAO Hubble data and the full 3D BAO signal crucially depend on a fiducial cosmological model, that assumes $c=c_0$, for extracting the peak signal. Hence, the r-BAO data contains systematics for the fiducial model without considering time-varying $c$. As for the transverse BAO measurements, they are nearly model-independent, being obtained through the 2-point angular correlation function of galaxies or QSOs applied on very narrow redshift ranges ($\delta z \sim 10^{-3}$). Nonetheless, there is minimal dependency on a fiducial cosmology when correct projection effects in those thin redshift bins, but it should not significantly affect the extraction of the angular-BAO signal, as explained in Refs. \cite{carvalho16,alcaniz17,carvalho20,decarvalho18,decarvalho21}. Again, the a-BAO data requires the assumption of a sound horizon scale for converting them into angular distance measurements. Although these factors impose a limitation in our consistency test, however, we do not ignore them. Our reconstruction is based on the combinations both including and excluding the BAO datasets.

%%%%%%%%%%%%%%%%%%%%%%%%%%%%%%%%%%%%%%%%%%%%%%%%%%%

\subsection{Reconstruction}

\begin{figure*}[t!]
	\begin{center}
		\includegraphics[angle=0, width=0.49\textwidth]{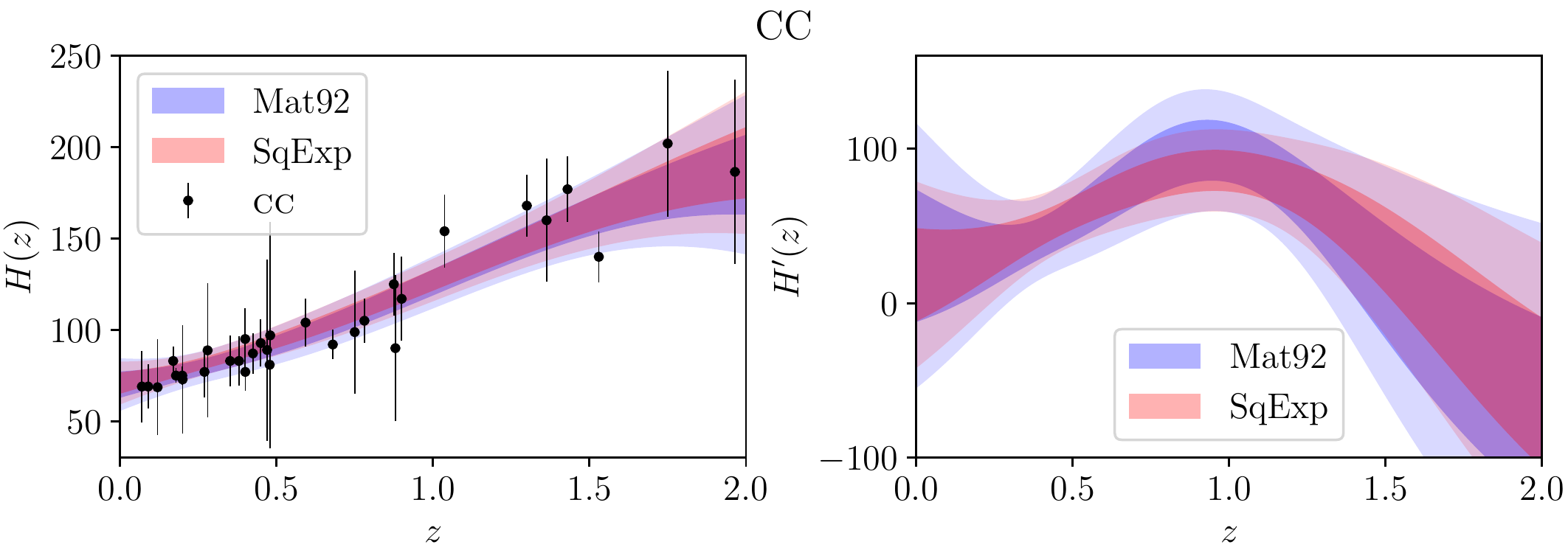}
		\includegraphics[angle=0, width=0.49\textwidth]{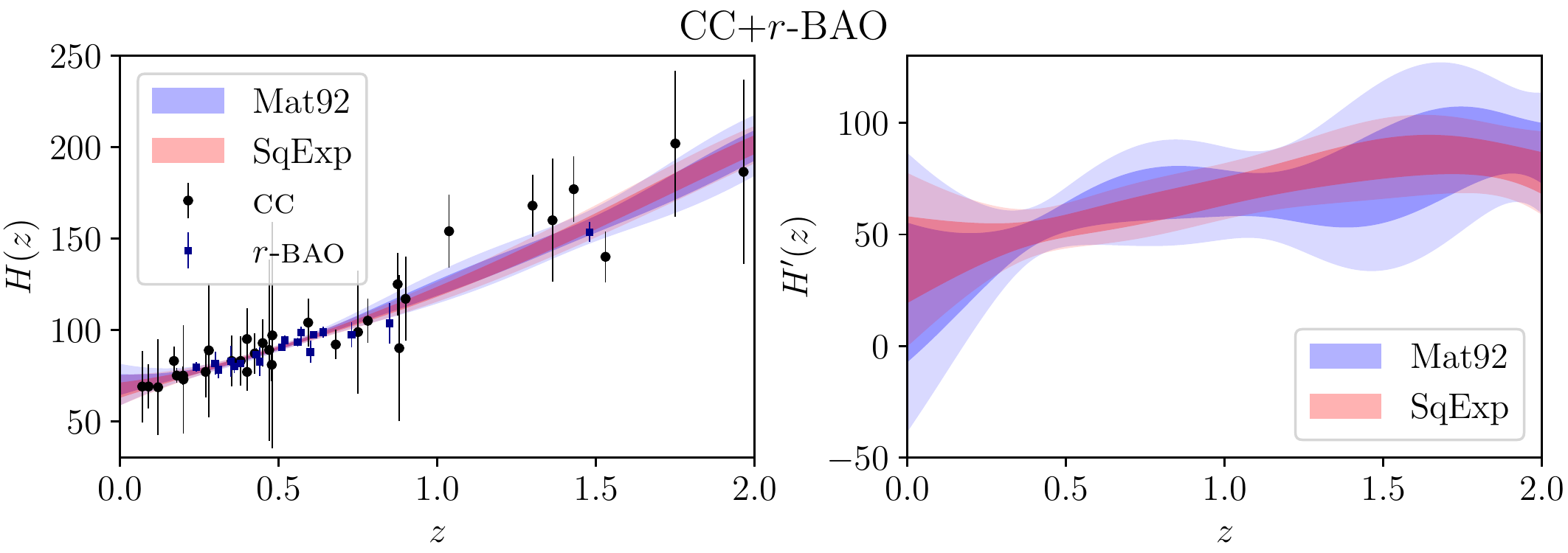}
	\end{center}
	\caption{{\small Plots for the reconstructed $H(z)$ with $H'(z)$ [in units of km Mpc$^{-1}$ s$^{-1}$] using CC (left) and CC+$r$-BAO (right) data. The shaded regions correspond to the associated 1$\sigma$ and 2$\sigma$ confidence levels.}} \label{fig:H-plot}
\end{figure*} 

\begin{figure*}[t!]
	\begin{center}
		\includegraphics[angle=0, width=\textwidth]{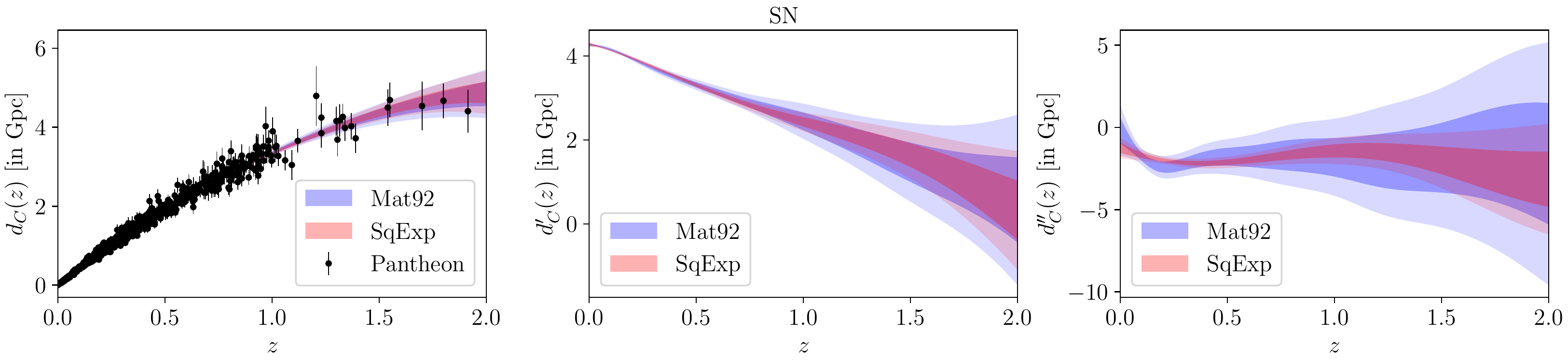}\\
		\includegraphics[angle=0, width=\textwidth]{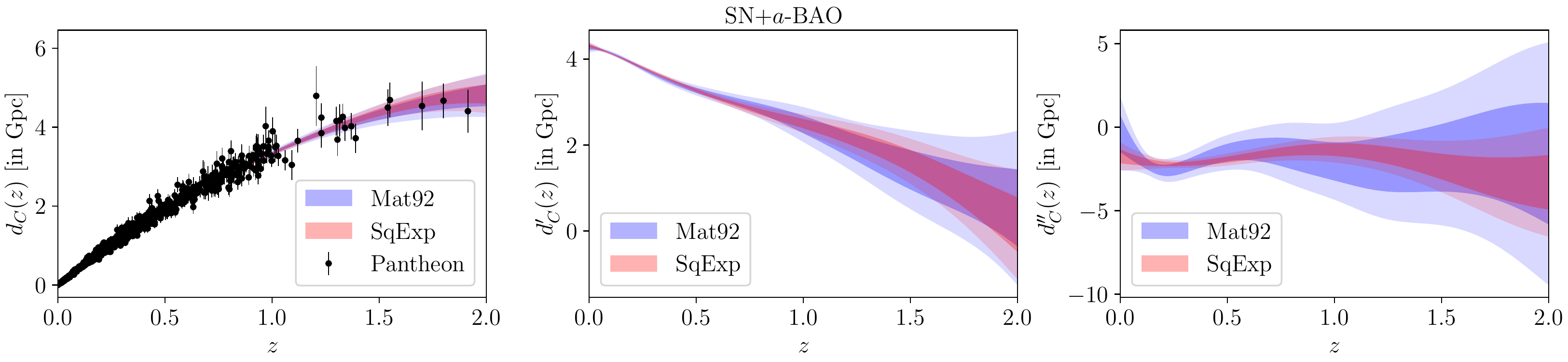}
	\end{center}
	\caption{{\small Plots for the reconstructed $d_C(z)$, $d_C'(z)$ with $d_C''(z)$ [in units of Gpc] using SN (top) data with $M_B = -19.382 \pm 0.054$ and SN+$a$-BAO (bottom) data with $M_B = -19.375 \pm 0.026$, as given in Table \ref{tab:MB_calib}. The shaded regions correspond to the associated 1$\sigma$ and 2$\sigma$ confidence levels.}} \label{fig:d_C-plot}
\end{figure*} 

We reconstruct the $H(z)$ and its derivative $H'(z)$ employing the non-parametric Gaussian Process regression with the {\sc GaPP} package~\cite{seikel12} (see also~\cite{shafieloo12}) on the Hubble parameter measurements from CCs \& CC+$r$BAOs. We adopt the squared exponential (hereafter referred to as SqExp) covariance, 
\begin{equation} \label{eq:sqexp}
k(z, \tilde{z}) = \sigma_f^2 \exp \left\lbrace - \frac{(z-\tilde{z})^2}{2l^2}\right\rbrace,
\end{equation}
and Mat\'{e}rn 9/2 (hereafter referred to as Mat92) covariance, 
\begin{equation} \label{eq:mat92}
k(z,\tilde{z}) = \sigma_f^2 \exp \left( \frac{-3 \vert z - \tilde{z} \vert}{l} \right) \left[ 1 + \frac{3 \vert z - \tilde{z} \vert}{l} + \frac{27 ( z - \tilde{z})^2}{7l^2} + \frac{18 \vert z - \tilde{z} \vert ^3}{7l^3} + \frac{27 \left( z - \tilde{z} \right)^4}{35l^4}\right] 
\end{equation}
functions as our default choices throughout the analysis. The SqExp kernel is indefinitely differentiable whereas the Mat92 is differentiable up to $3$rd order. The present work requires the use of second-order derivatives, thus, we can compare the results obtained for the choice of these two kernels. We assume $N = 1000$ reconstruction bins along redshift range $0 < z < 2$. Following the prescription given by Rodrigues \& Bengaly~\cite{gabriel22}, we do not optimize the GP hyperparameters in order to avoid possible biases in these reconstructions due to overfitting. A similar result was also found in~\cite{eoin}. Instead of fixing the hyperparameters ($l$, $\sigma_f$) to their best-fit values, we consider representative samples from their posteriors that lie within the 1$\sigma$ confidence region and finally combine the individual predictions to obtain the final reconstruction.

Plots for the reconstructed $H(z)$ and $H'(z)$ for the CC Hubble and CC+$r$-BAO Hubble data are shown in Fig. \ref{fig:H-plot}. Here, we bring to notice that the CC and joint CC+$r$-BAO data span up to redshift $z = 1.965$ and $2.36$ respectively, where $r$-BAOs provide data at higher $z$ with better precision than CC. This causes a significant dip in the reconstruction profile of $H^\prime(z)$ when working with CC instead of the CC+$r$-BAO combination.

Further, we derive the transverse comoving distances $d_C$ from the Pantheon SN data as,
\begin{equation} \label{eq:d_C-SN}
    {d_C}^{\text{SN}} = \frac{d_L}{1+z} = \frac{1}{1+z}10^{\frac{m-M_B}{5}-5},
\end{equation} where $M_B$ is the absolute magnitude of SN-Ia. For a self-consistent analysis, as the absolute magnitude $M_B$ of SN-Ia is degenerate with the value of $H_0$, we need to obtain the marginalized constraints on $M_B$. We also make use of the transverse BAO measurements, as 
\begin{equation}\label{eq:d_C-BAO}
    {d_C}^{\text{BAO}} = d_A (1+z) = \frac{r_d}{\theta_{\text{BAO}}} (1+z) , 
\end{equation}
where $r_d$ is the comoving sound horizon at the drag epoch. We undertake a GP reconstruction of $d_C(z)$, as well as its higher derivatives $d_C'(z)$ and $d_C''(z)$, substituting the value of $M_B$ and $r_d$ in Eq. \eqref{eq:d_C-SN} and Eq. \eqref{eq:d_C-BAO}. 

Therefore, our key equation for the reconstruction of $T(z)$, given in Eq \eqref{eq:Tz}, reduces to,
\begin{widetext}
	\begin{equation}\label{eq:Tz-recon}
	T(z) = 1 + c_0^{-2}\left[ \left\lbrace d_C''(z)H(z) + d_C'(z)H'(z) \right\rbrace d_C(z)H(z) - {d_C'^2}(z){H^2}(z) \right].
	\end{equation}
\end{widetext}
We consider $c_0 = 3 \times 10^{5}$ km s$^{-1}$ and make use of Eqs. \eqref{eq:Hubble_reduced} and \eqref{eq:dc_vs_D} that links our reconstructed functions ($d_C(z)$, $H(z)$ and its derivatives) to their normalized counterparts, $D(z)$ and $E(z)$ respectively. Also note that we have implicitly assumed the validity of the cosmic distance duality relation (CDDR) by means of Eqs.~\eqref{eq:d_C-SN} and~\eqref{eq:d_C-BAO}. Thus, our consistency test will be able to distinguish between the SCM and some classes of VSL models - e.g. the so-called mVSL models~\cite{barrow98}, as long as they satisfy the CDDR, which is not the case for the meVSL models~\cite{lee21a}. For the sake of completeness and to check the consistency of the results, we also examine the effect of different $H_0$ priors and $M_B$ priors on the reconstruction of $T(z)$. This exercise can help us test the mutual consistency of the datasets in use, as well as understand if there exists any hidden systematics in the data. In view of the rising tensions in cosmology, we have adopted a rigorous analysis to shed more light on this intriguing puzzle. 

\squeezetable
\begin{table*}[t!]
	\caption{{\small Table showing the marginalized constraints on $M_B$ (i) assuming a fiducial $\Lambda$CDM model; (ii) with GP following a similar prescription described in Ref. \cite{purba_j,purba_q}.}}
	\begin{center}
		\rule{0.98\textwidth}{1pt} 
		
		\resizebox{\textwidth}{!}{\renewcommand{\arraystretch}{1.5} \setlength{\tabcolsep}{25 pt} \centering  
			\begin{tabular}{l  c  c  c }
				
				\textbf{Datasets} &  Method (i) & \multicolumn{2}{c}{Method (ii)} \\ 
				\hline
				& 	$\Lambda$CDM & SqExp & Mat92 \\
				\hline
				CC+SN & $-19.382 \pm 0.054$	& $-19.254 \pm 0.056$ & $-19.361 \pm 0.059$ \\		
				CC+$r$-BAO+SN & $-19.375 \pm 0.026$ & $-19.357 \pm 0.036$ & $-19.382 \pm 0.041$ \\
			\end{tabular} 
		}
		\rule{0.98\textwidth}{1pt}
	\end{center}
	\label{tab:MB_calib}
\end{table*}

\begin{figure*}[t!]
	\begin{center}
		\includegraphics[angle=0, width=0.3\textwidth]{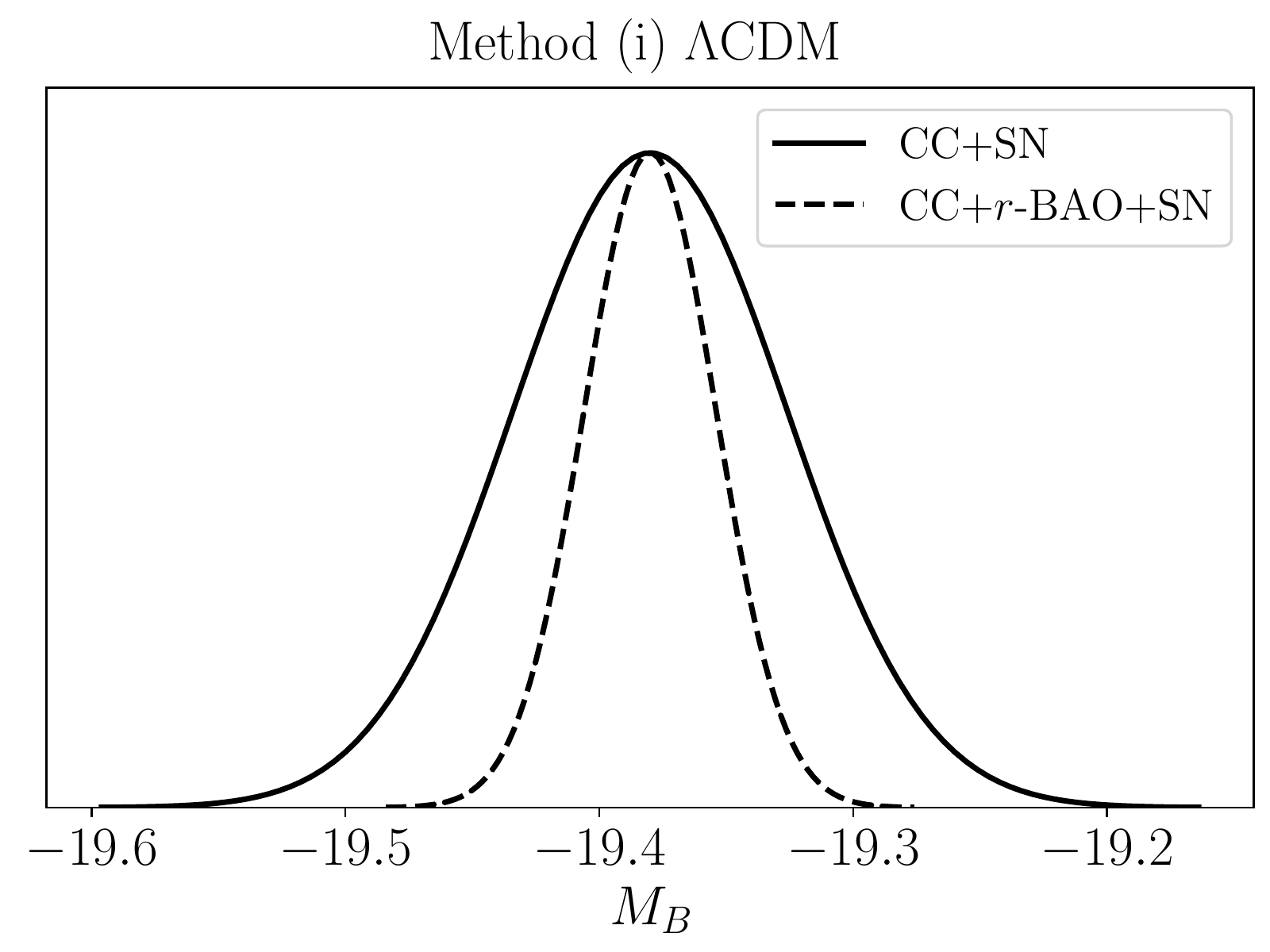} \hfill 
		\includegraphics[angle=0, width=0.3\textwidth]{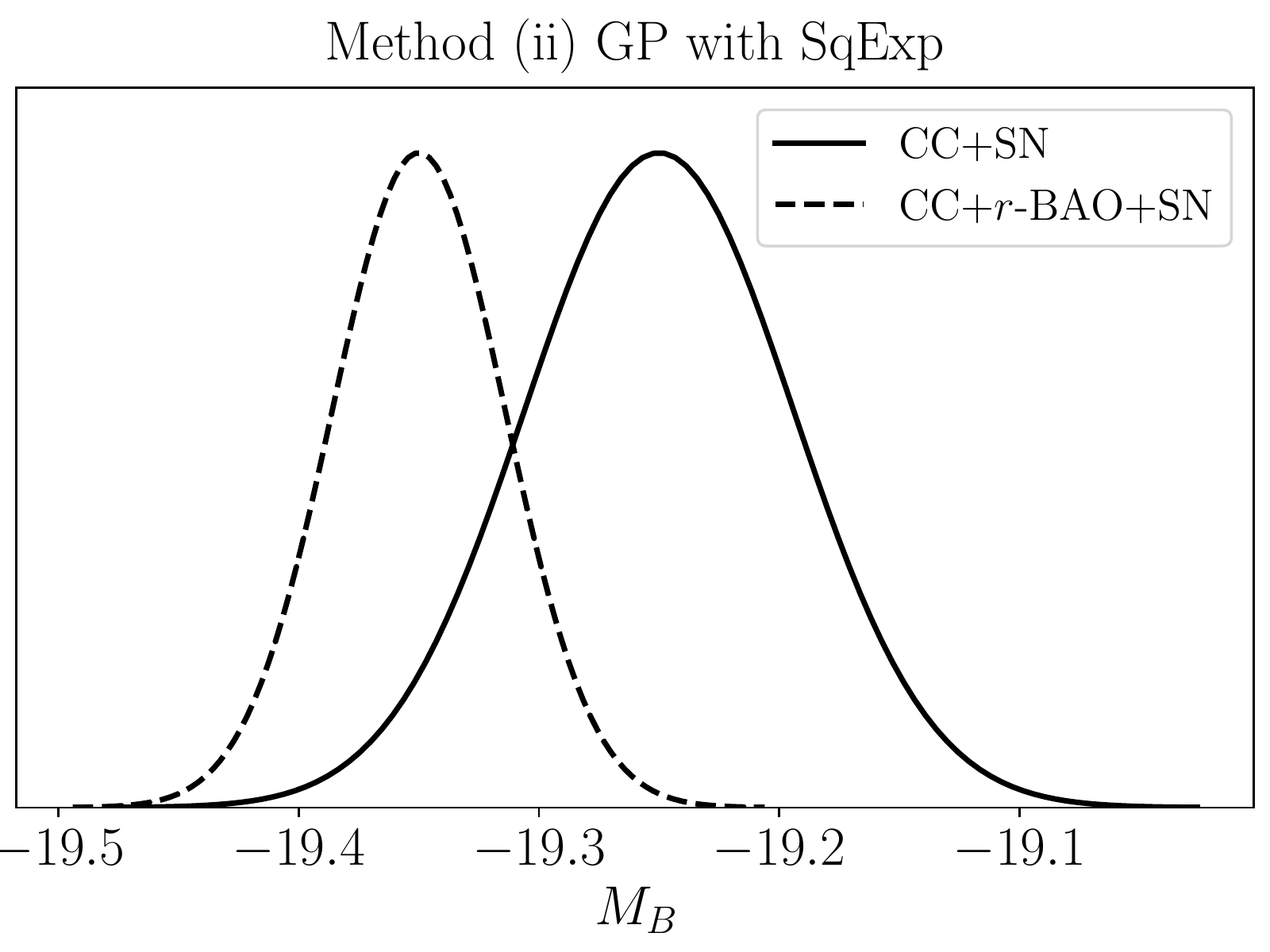} \hfill 
		\includegraphics[angle=0, width=0.3\textwidth]{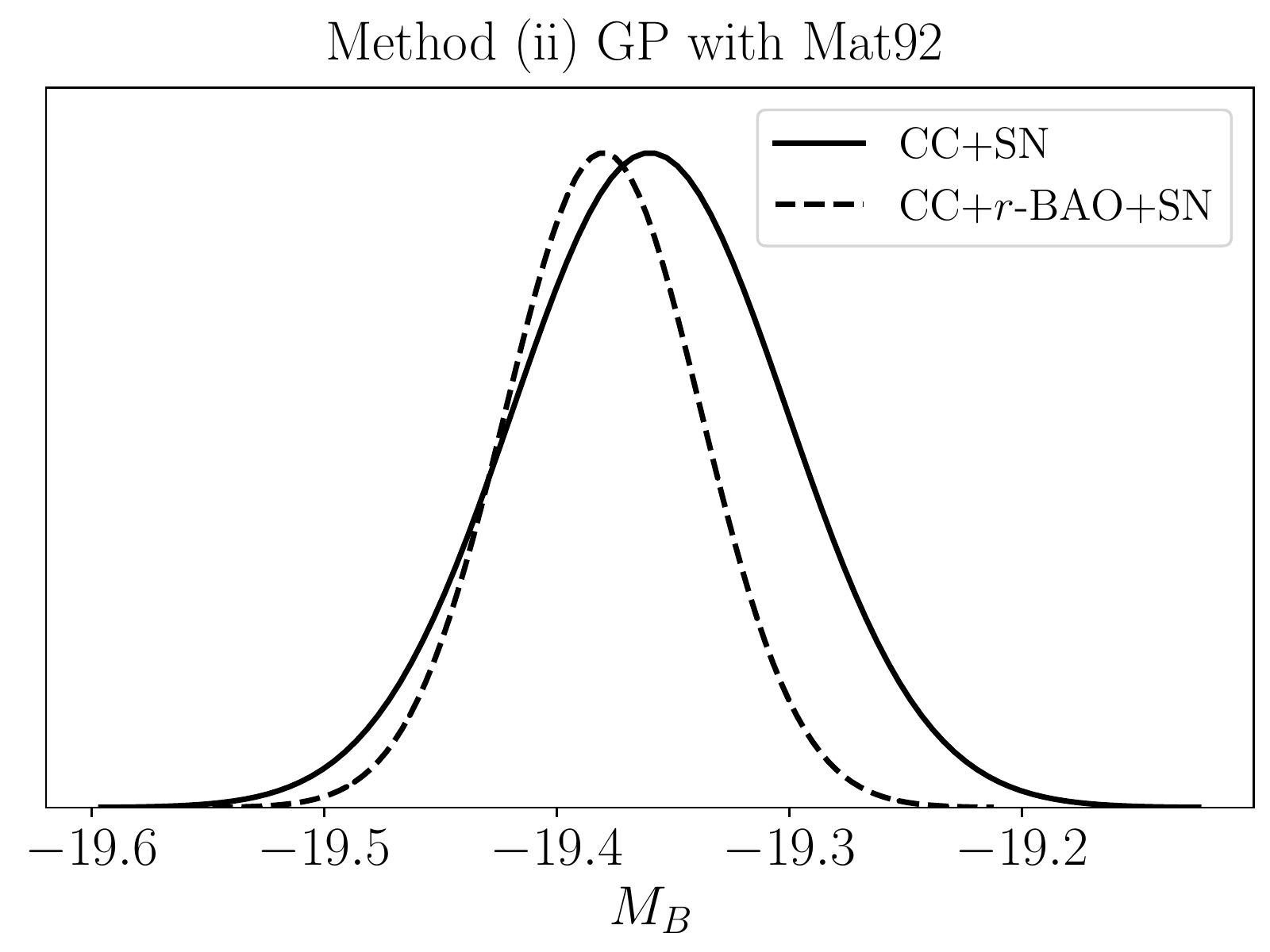} 
	\end{center}
	\caption{{\small Plots showing likelihood for the calibrated values of $M_B$ for Method (i) assuming a fiducial $\Lambda$CDM model (left); (ii) with GP with the Squared Exponential [SqExp] kernel (centre) and Matern 9/2 [Mat92] kernel (right)}}. \label{fig:MB_calib}
\end{figure*}

%%%%%%%%%%%%%%%%%%%%%%%%%%%%%%%%%%%%%%%%%%%%%

\section{Results}

\begin{figure*}[t!]
	\begin{center}
		\includegraphics[angle=0, width=\textwidth]{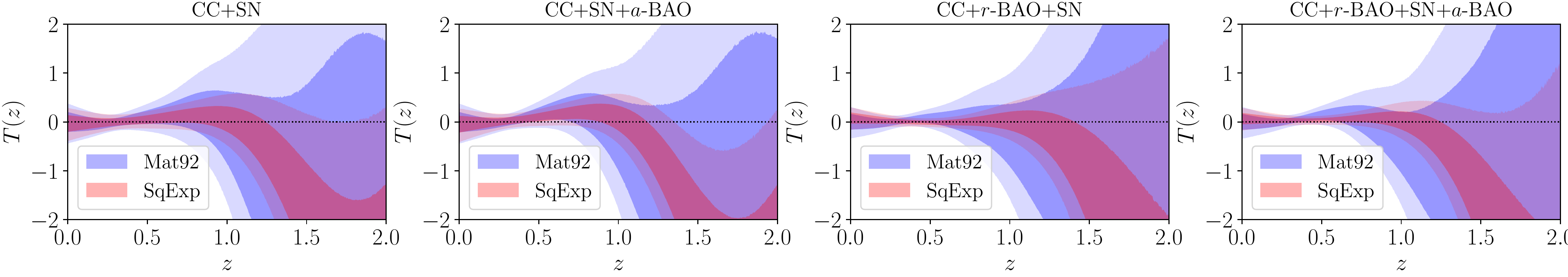}\\
		\includegraphics[angle=0, width=\textwidth]{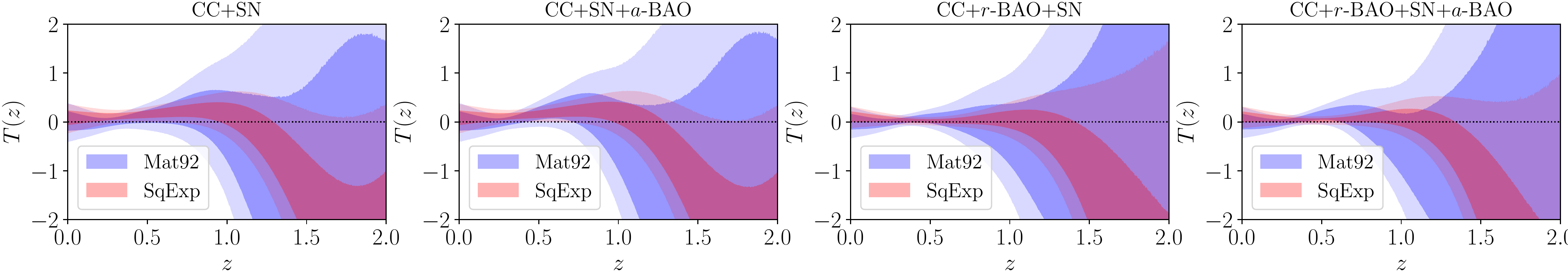}
	\end{center}
	\caption{{\small Plots for the reconstructed $T(z)$ using the calibrated values of $M_B$ from Method (i) assuming a fiducial $\Lambda$CDM model (top row); (ii) with model-independent GP (bottom row) for different combinations of datasets. The shaded regions correspond to the associated 1$\sigma$ and 2$\sigma$ confidence levels.}} \label{fig:Tz_plot}
\end{figure*}

\begin{table*}[t!]
	\caption{{\small Table showing the calibrated values of $M_B$ using the (i) SH0ES; (ii) TRGB; and (iii) Planck; $H_0$ value as priors.}}
	\begin{center}
		\rule{0.98\textwidth}{1pt} 
		\resizebox{\textwidth}{!}{\renewcommand{\arraystretch}{1.5} \setlength{\tabcolsep}{25 pt} \centering  
			\begin{tabular}{l  c  c  c }
				
				\textbf{Datasets} &  \multicolumn{3}{c}{$H_0$ prior} \\ 
				\hline 
				&  (i) SH0ES & (ii) TRGB &  (iii) Planck\\  
				\hline
				CC+$H_0$+SN & $-19.302 \pm 0.031$	& $-19.369 \pm 0.036$ & $-19.427 \pm 0.020$ \\		
				CC+$H_0$+$r$-BAO+SN & $-19.340 \pm 0.020$ & $-19.372 \pm 0.022$ & $-19.417 \pm 0.015$ \\
			\end{tabular} 
		}
		\rule{0.98\textwidth}{1pt}\\
	\end{center}
	\label{tab:MB_H0_calib}
\end{table*}

\begin{figure*}[t!]
	\begin{center}
		\includegraphics[angle=0, width=\textwidth]{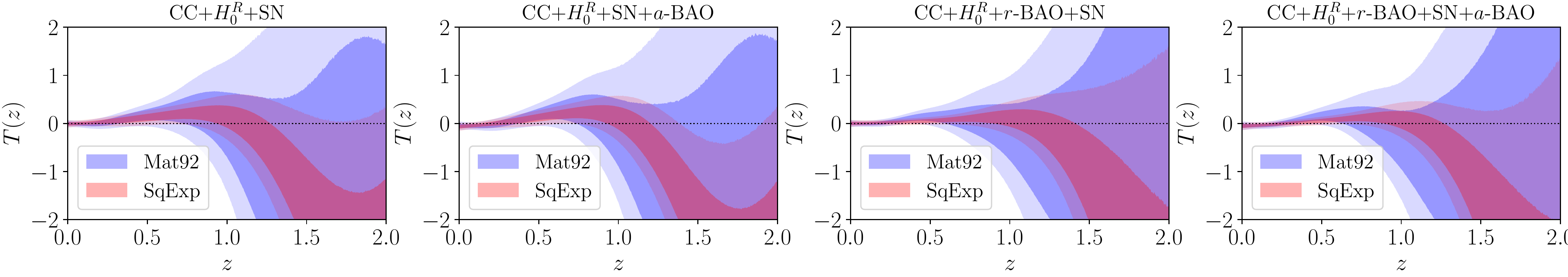}\\
		\includegraphics[angle=0, width=\textwidth]{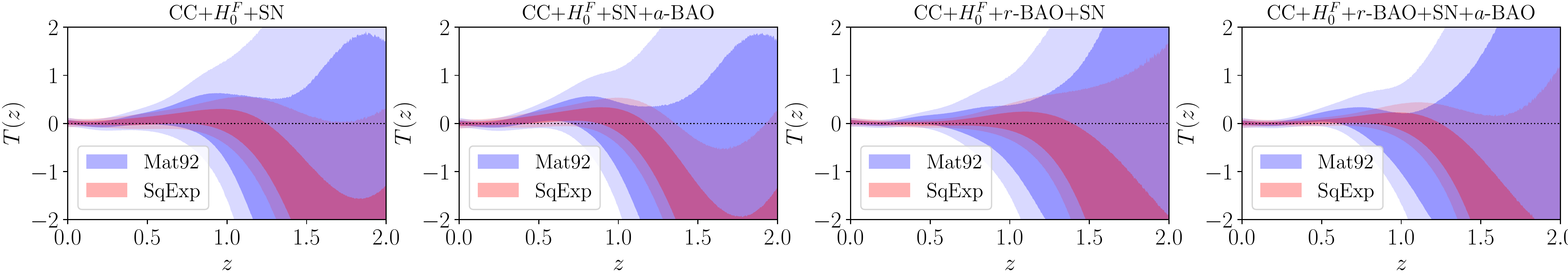}\\
		\includegraphics[angle=0, width=\textwidth]{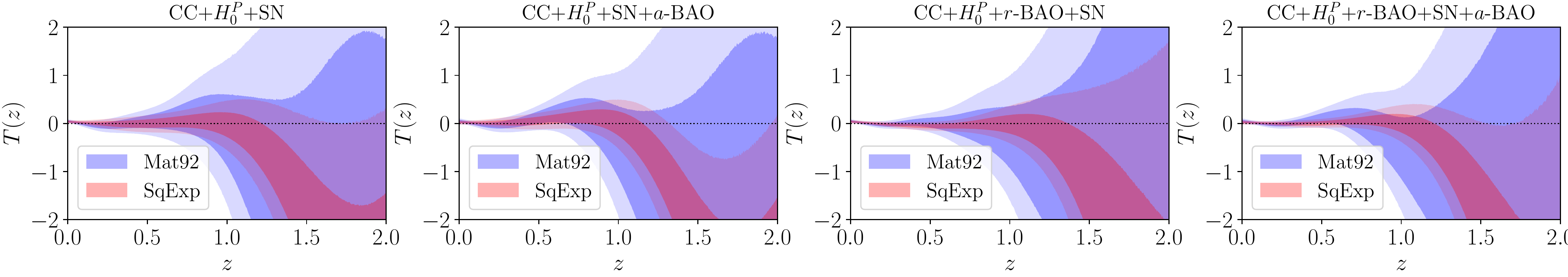}
	\end{center}
	\caption{{\small Plots for the reconstructed $T(z)$ using the (i) SH0ES (top row); (ii) TRGB (middle row); and (iii) Planck (bottom row); $H_0$ value as priors. The shaded regions correspond to the associated 1$\sigma$ and 2$\sigma$ confidence levels.}} \label{fig:Tz_H0_plot}
\end{figure*}

\begin{figure*}[t!]
	\begin{center}
		\includegraphics[angle=0, width=\textwidth]{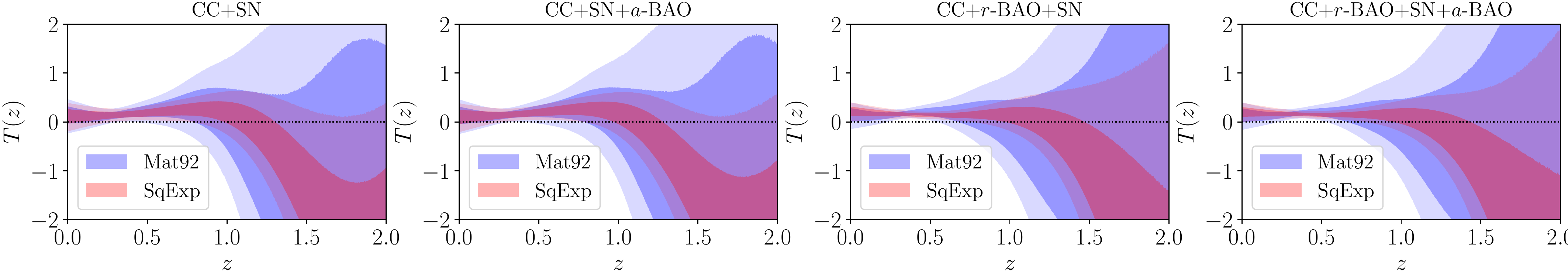}\\
		\includegraphics[angle=0, width=\textwidth]{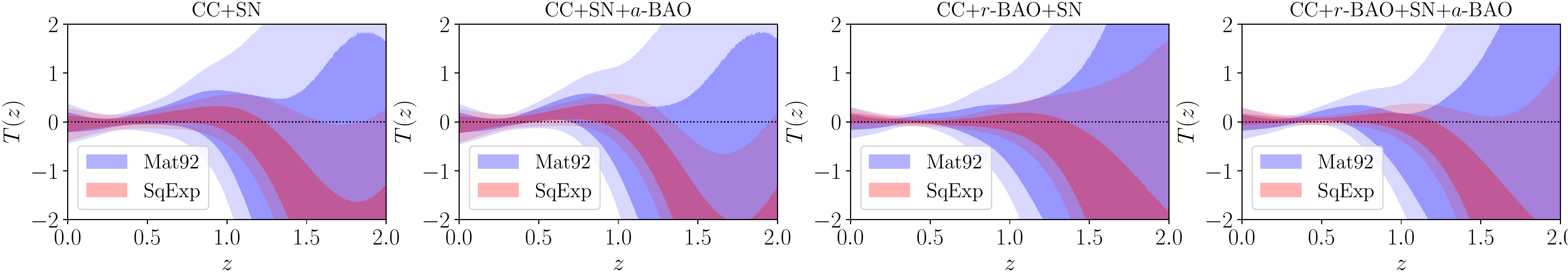}\\
		\includegraphics[angle=0, width=\textwidth]{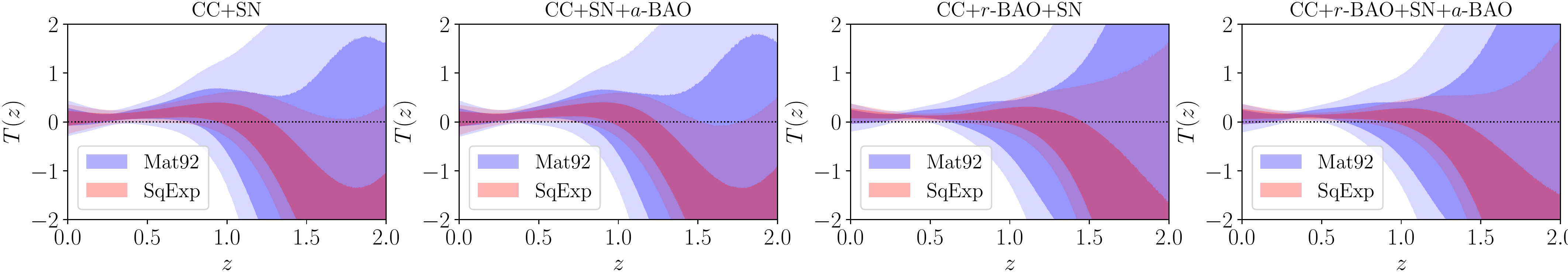}\\
		\includegraphics[angle=0, width=\textwidth]{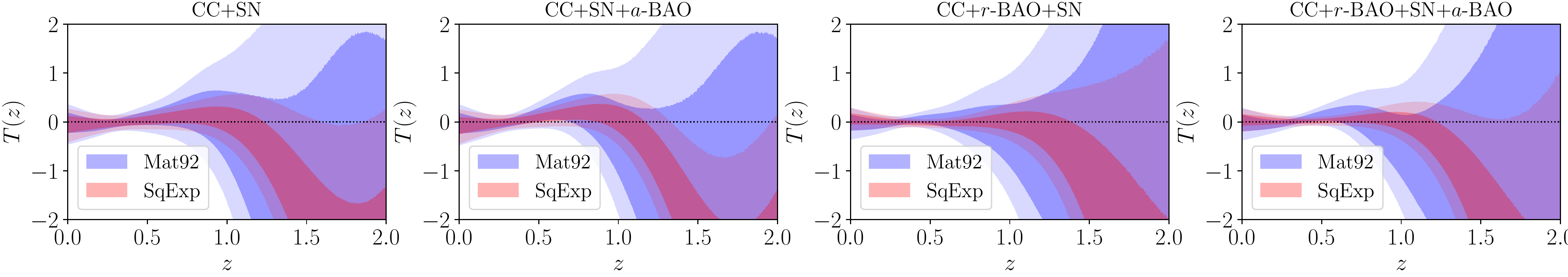}
	\end{center}
	\caption{{\small Plots for the reconstructed $T(z)$ using four different priors on $M_B$ from existing literature: (i) $M_B = -19.214 \pm 0.037$ given by Riess {\it et al}\cite{riess22} (first row); (b) $M_B = -19.387 \pm 0.021$ by Efstathiou\cite{efstathiou21} (second row); (c) $M_B = -19.401 \pm 0.027$ from anisotropic-BAO (third row); and (d) $M_B = -19.262 \pm 0.030$ from angular-BAO measurements (fourth row), by Camarena \& Marra\cite{camarena21}, respectively. The shaded regions correspond to the associated 1$\sigma$ and 2$\sigma$ confidence levels.}} \label{fig:Tz_MB_plot}
\end{figure*}

To start with, we assume the vanilla $\Lambda$CDM model to obtain the constraints on $M_B$. For the transverse angular-BAO data, consider $r_d = 147.09 \pm 0.26$ Mpc obtained from the Planck \cite{planck21} (TT,TE,EE+lowE+lensing 2018) result.  The plots for the reconstructed $d_C(z)$, $d_C'(z)$ and $d_C''(z)$ from the SN and SN+$a$-BAO data are shown in Fig. \ref{fig:d_C-plot}. Furthermore, to reduce the model-dependence in our results, we follow a similar prescription as in Ref. \cite{purba_j,purba_q} to obtain the marginalized constraints on $M_B$ in a more robust manner. Plots for the marginalized $M_B$ constraints are shown in Fig. \ref{fig:MB_calib}. The best-fit values with the associated 1$\sigma$ uncertainties are given in Table \ref{tab:MB_calib}. Finally, with reconstructed $d_C(z)$, $d_C'(z)$ and $d_C''(z)$ we derive $T(z)$ using Eq. \eqref{eq:Tz-recon}. The reconstructed $T(z)$ using various combinations of data sets for the two choices of covariance function are shown in Fig. \ref{fig:Tz_plot}. The shaded regions correspond to the 68\% and 95\% C.L., respectively, from darker to lighter shades. We find that $T = 0$ is mostly consistent within 2$\sigma$ throughout the redshift range $0<z<2$.

As a general comment, we highlight that the reconstruction kernel plays a significant role and the final results are sensitive to this choice. By construction, the SqExp covariance function is infinitely differentiable, hence it has an infinite degree of smoothness. This assumption leads to strong correlations in the reconstructed functions $H(z)$, $H^\prime(z)$, $d_c(z)$, $d_c^\prime(z)$ and $d_c^{\prime\prime}(z)$, which gives smaller bounds on the uncertainties as shown in Fig. \ref{fig:H-plot} and \ref{fig:d_C-plot} in comparison to the Mat92 kernel.

\begin{figure*}[t!]
	\begin{center}
		\includegraphics[angle=0, width=\textwidth]{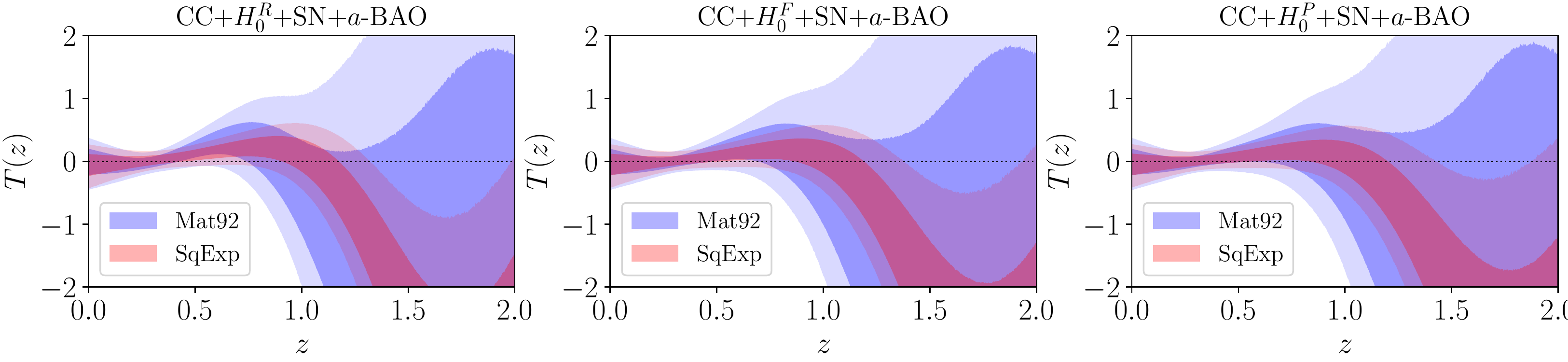}\\
		\includegraphics[angle=0, width=\textwidth]{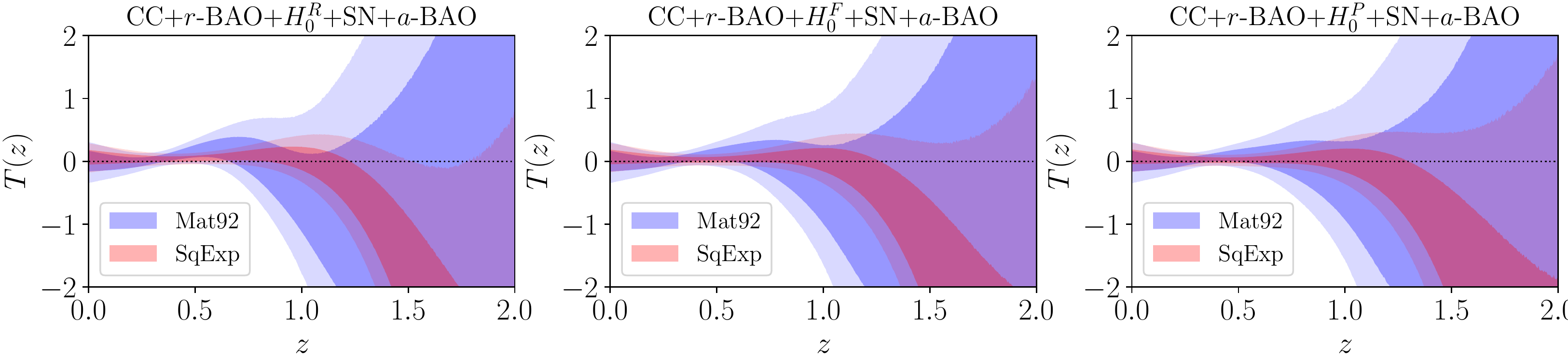}\\
	\end{center}
	\caption{{\small Plots for the reconstructed $T(z)$ using $r_d h = 102.56 \pm 1.87$, considering the SH0ES (left column), TRGB (middle column) and Planck (right column) $H_0$ priors. The shaded regions correspond to the associated 1$\sigma$ and 2$\sigma$ confidence levels.}} \label{fig:Tz_rd_plot1}
\end{figure*}

\begin{figure*}[t!]
	\begin{center}
				\includegraphics[angle=0, width=\textwidth]{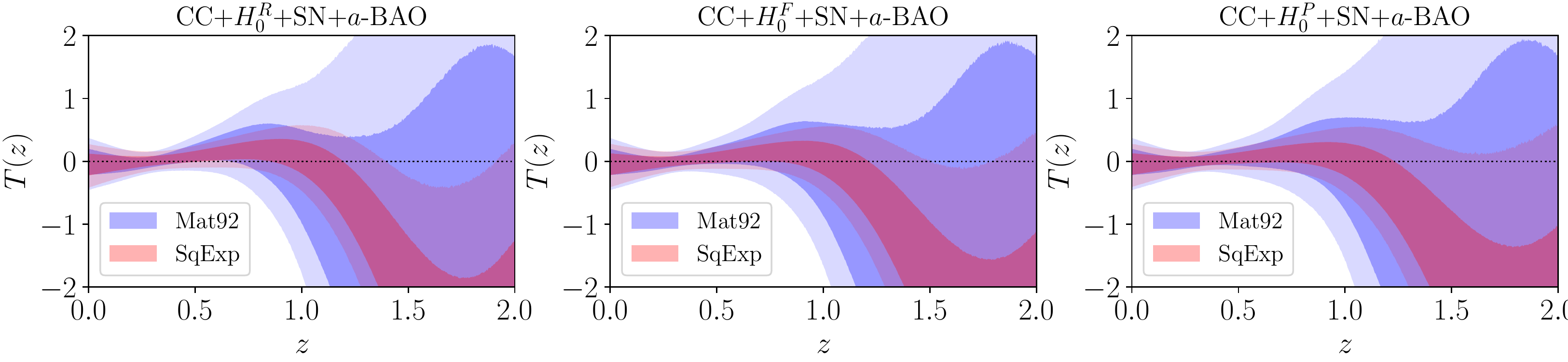}\\
				\includegraphics[angle=0, width=\textwidth]{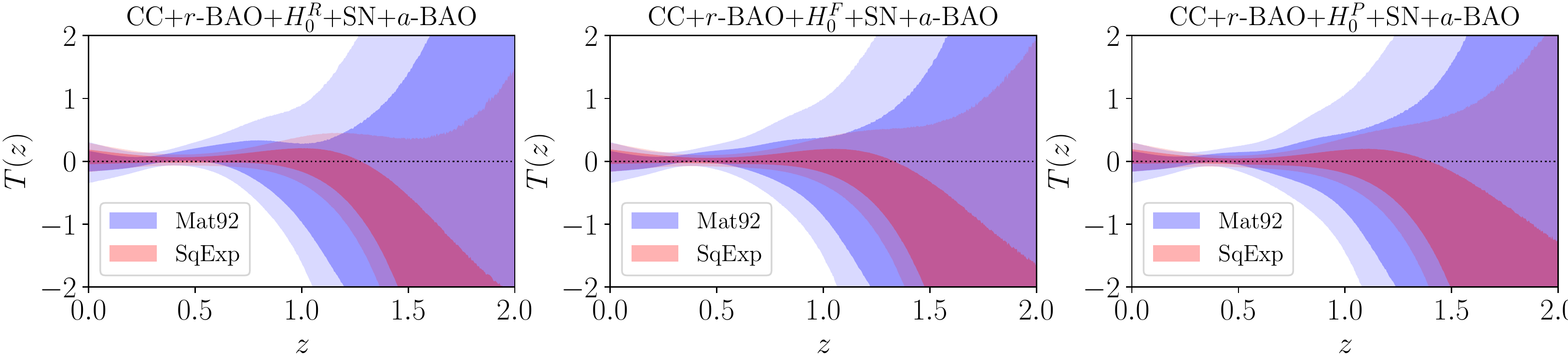}\\
	\end{center}
	\caption{{\small Plots for the reconstructed $T(z)$ using $r_d h = 109.37 \pm 2.09$, considering the SH0ES (left column), TRGB (middle column) and Planck (right column) $H_0$ priors. The shaded regions correspond to the associated 1$\sigma$ and 2$\sigma$ confidence levels.}} \label{fig:Tz_rd_plot2}
\end{figure*}

Different strategies for determining the present value of the Hubble parameter are well-known in the literature. We have taken into account three values $H_0$, namely (i) the local measurements by the SH0ES team ($H_0^{\text{R}}$ = $73.2 \pm 1.3$ km Mpc$^{-1}$ s$^{-1}$ \cite{riess21}), (ii) the updated TRGB calibration from the Carnegie Supernova Project ($H_0^{\text{F}}$ = $69.8 \pm 1.7$ km Mpc$^{-1}$ s$^{-1}$ \cite{freedman21}), and (iii) the inferred value from CMB sky via extrapolation of data on the early universe by the Planck survey ($H_0^{\text{P}}$ = $67.27 \pm 0.6$ km Mpc$^{-1}$ s$^{-1}$ \cite{planck21}), respectively, to examine the present status of the Hubble tension and its effect on the reconstruction.  When considering the effect of $H_0$ priors, we again need to calibrate the absolute magnitude $M_B$ keeping in mind the degeneracy between $H_0$ and $M_B$. So, for this exercise, we assume the vanilla $\Lambda$CDM model and obtain the constraints on $M_B$, given in Table \ref{tab:MB_H0_calib}. Finally, we reconstruct $T(z)$ corresponding to different values of $H_0$ in Fig. \ref{fig:Tz_H0_plot}.

The constancy of SN-Ia absolute magnitude $M_B$ has often been questioned as an immediate retort to the $H_0$ tension, with a major part of the scientific community seeking some unknown systematics. In recent years, several studies have raised the question of whether or not $M_B$ might evolve with redshift. The principal reason being the value of $M_B$, which is used to derive the local $H_0$ constraint, is not compatible with that necessary to fit the BAO and CMB data. The SH0ES team measures $M_B$, by calibrating the distances of SN-Ia host galaxies to local geometric distance anchors via the Cepheid luminosity relation. This $M_B$ is then converted into $H_0$ via the magnitude-redshift relation of the Pantheon compilation. Therefore, the tension in $H_0$ has often been argued as a tension associated with $M_B$ as the SH0ES $H_0$ measurement comes directly from the estimated $M_B$. The CMB constraint on the sound horizon using an inverse distance ladder predicts $M_B \sim -19.4$, while the SH0ES estimate corresponds to $M_B \sim -19.2$. Hence, we also focus on whether or not this tension in $M_B$ affects the constancy of the speed of light. In this work, have considered four priors on $M_B$, namely (a) $M_B = -19.214 \pm 0.037$ given by Riess {\it et al}\cite{riess22}; (b) $M_B = -19.387 \pm 0.021$ from the inverse distance ladder by Efstathiou\cite{efstathiou21}; (c) $M_B = -19.401 \pm 0.027$ from anisotropic-BAO; and (d) $M_B = -19.262 \pm 0.030$ from angular-BAO measurements both by Camarena \& Marra\cite{camarena21}. Plots for the reconstructed $T(z)$ corresponding to different values of $M_B$ are shown in Fig. \ref{fig:Tz_MB_plot}.

We further examine the effect of the sound horizon scale on the reconstruction when using the BAO data. For this exercise, we consider two constraints on $r_d h$ obtained by Camarena \& Marra\cite{camarena20}, namely (i) $r_d h = 102.56 \pm 1.87$, and (ii) $r_d h = 109.37 \pm 2.09$, where $h \equiv \frac{H_0}{100 \text{ km Mpc$^{-1}$ s$^{-1}$}}$ is the dimensionless Hubble parameter. We take into account the three values of $H_0$, namely the Planck, SH0ES and TRGB priors as discussed earlier, to compute $r_d$. These $r_d$ estimates are then utilized to obtain the comoving distance measurements ${d_C}^{\text{BAO}}$ from the $a$-BAO data and carry out the reconstruction of $T(z)$. Finally, we plot the reconstructed $T(z)$ for both the cases (i) $r_d h = 102.56 \pm 1.87$ and (ii) $r_d h = 109.37 \pm 2.09$, in Figs. \ref{fig:Tz_rd_plot1} and \ref{fig:Tz_rd_plot2}, respectively.\\

In a nutshell, our key findings can be summarized explicitly as follows: \begin{enumerate}
	\item We find null evidence for the speed of light variability hypothesis for most choices of priors and data-set combinations (Fig. \ref{fig:Tz_plot}). The final results are sensitive to the choice of the reconstruction kernel. Imposing greater differentiability, hence a greater degree of smoothness leads to tighter bounds on the uncertainties. 
    
        \item Mild deviations at $2\sigma$ confidence level are seen for $z>1$ with the SqExp covariance function when the CC+SN+$a$-BAO data is employed (row 1, column 2 of Fig. \ref{fig:Tz_plot}). This discrepancy could arise from two factors. Firstly, due to the incompleteness of the $a$-BAO data set at higher redshifts. Secondly, the choice of this infinitely differentiable kernel seems to result in strong correlations and smaller errors in the reconstructions. 
        
	\item Inclusion of $H_0$ priors imposes tighter constraints on this consistency test at lower redshifts i.e. $z < 0.5$ (Fig. \ref{fig:Tz_H0_plot}). As $z$ increases beyond $z>1$, we observe departure from $T=0$ for the CC+$H_0$+SN+$a$-BAO combinations with the SqExp kernel. However, we ascribe no statistical significance to this result for the aforementioned reason.
	
	\item We observe that $T=0$ is almost included in 2$\sigma$ for the CC+SN combinations (column 1 in Fig \ref{fig:Tz_MB_plot}), independent of the choice of $M_B$ priors. Mild deviations are visible on employing the SqExp covariance for $M_B \sim -19.2$ at $z<1$ and for $M_B \sim -19.4$ at $z>1$. However, this could again be an artefact of the kernel itself. 
 
        \item Inclusion of the $r$-BAO data set (column 3 in Fig \ref{fig:Tz_MB_plot}) leads to regions where $T \neq 0 $ at $z< 1$ for $M_B \sim -19.2$ for both the kernel choices, but there is no violation for $M_B \sim -19.4$ cases. This departure clearly arises due to the choice of inconsistent priors, $M_B$ and $r_d$, as both are degenerate with $H_0$. 
        
        \item There are no marked differences when the $a$-BAO data is used in combination with the CC, SN and $r$-BAO data sets, except for a mild dip at $z>1$ for $M_B \sim -19.4$ with the SqExp kernel (column 2 and 4 in Fig. \ref{fig:Tz_MB_plot}).
 
	\item Figs. \ref{fig:Tz_rd_plot1} and \ref{fig:Tz_rd_plot2} indicate the $a$-BAO data prefers a higher value of $r_d h$ and lower value of $H_0$, which is obvious for the SqExp kernel. This leads to an estimate of $r_d$ quite different from that of the Planck prediction.

\end{enumerate}

%%%%%%%%%%%%%%%%%%%%%%%%%%%%%%%%%%%%%%%%%%%%%%%%%%%

\section{Discussion \label{conclusion}}

Testing the variability of fundamental constants in Nature is one of the strongest tests of fundamental Physics. Any significant evolution of their values would immediately hint at new physics, and demand a profound reformulation of the standard model of Cosmology and Particles, not to mention Electromagnetism, Thermodynamics and Gravitation. Although these tests have been extensively performed on Earth labs and in our Solar neighbourhood with tremendous precision, reporting null evidence for their evolution, there are not many cosmological tests of this kind as yet, due to the limitations of high redshift observations. Hence, there is an urge to pursue these tests as a means to challenge the validity of the standard cosmological model - particularly in view of its theoretical and observational challenges.  

Under this motivation, we revisit the consistency test of the speed of light variability proposed by~\cite{cai16} using the latest cosmological observations at the $0<z<2$ redshift range. Conversely from previous analyses~\cite{salzano15, gabriel22}, this test circumvents the assumption of a flat Universe, which is important since a non-flat cosmic curvature can be degenerated with an evolving speed of light, as shown by~\cite{salzano17}. On the other hand, this test is more computationally extensive, as we need to compute higher-order derivatives of cosmological distances and ages. We deploy Gaussian Processes in order to reconstruct the cosmological information from differential galaxy ages (Cosmic Chronometers), Type Ia Supernova apparent magnitudes, and measurements of both radial and transverse Baryonic Acoustic Oscillations, in a model-independent way. We also explore the possible dependency of prior assumptions in this test, such as the Supernova absolute magnitude, the Hubble constant value, as well as the sound horizon scale. 

We find no significant hints at speed of light variability from CC, SN and radial BAO data, regardless of the cosmological priors and reconstruction kernel adopted, except for a mild deviation from this hypothesis ($\sim 2\sigma$ confidence level at $z>1.0$) when the transverse BAO measurements are included. Such a result is found to be moderately dependent on cosmological priors, as some assumptions of $M_B$ and $r_d$ do not lead to this deviation, but the strongest dependency comes from reconstruction kernel choice, as we only find this trend when the squared exponential kernel is assumed. However, we must note that this specific data set is very sparse on the redshift range where this result was found, and hence we ascribe it to its statistical limitation rather than a potential departure from the standard cosmological model. For all other data-set combinations, different prior choices have little impact on the results. Nevertheless, we note that the inclusion of $H_0$ priors imposes tighter constraints on this consistency test at lower redshifts ($z < 0.5$), while $r_d$ priors impose stronger bounds at the higher redshift ranges. Moreover, our null result hints towards the viability of the second law of thermodynamics for homogeneous and isotropic universes, regardless of the signature of the spatial curvature \cite{pavon}.

We further observe that imposing greater differentiability in the GP kernel gives us tighter bounds on the uncertainties in the reconstructed plots. This feature is clearly visible when working with the squared exponential kernel, as demanding infinite differentiability, and hence a greater degree of smoothness seems to lead to stronger correlations and smaller errors in the reconstructions.  So, it is important to identify the optimal kernels when undertaking GP, by employing some kernel selection methods. Besides, one should be careful while assuming certain priors values in cosmology, as degeneracies associated with these parameters can give rise to inconsistent results.

As a note of caution, we remark that some observational measurements and priors herein used make the assumption of specific cosmological models (as in the case of the radial BAO data set), as well as fundamental astrophysical relations that depend on a combination of other fundamental constants (for example the $M_B$ priors from Type Ia Supernovae). But since we are only focused on performing a consistency test of the speed of light variability in regards to the cosmological information that is currently available, we will leave a careful examination on how we could circumvent these issues for future work. Still, we would like to stress that this null test would be able to distinguish between the standard model and a class of minimal varying speed of light models, as in the form $c(z) = c_0 \hat{c}(z)$, for any non-flat Universe, granted that the cosmic distance duality relation is maintained. A thorough evaluation on which alternative models could be more easily distinguished through this test will also be pursued as a follow-up work. 

As a final note, we plan to carry out a forecast of this null test performance in light of forthcoming (and ongoing) cosmological redshift surveys, such as J-PAS, Euclid, SKAO, and LSST~\cite{jpas14, euclid18, lsst18, ska20}, along with next-generation gravitational wave observatories~\cite{punturo10}. We expect they will be able to enormously improve its precision.

\section*{Acknowledgements}

The authors would like to thank the anonymous referees for their constructive suggestions and valuable comments that led to a definite improvement in the work. PM thanks ISI Kolkata for financial support through Research Associateship. GR and CB acknowledge financial support by CAPES and FAPERJ postdoc recente nota 10 fellowships (PDR10), respectively.

%\newpage

%%%%%%%%%%%%%%%%%%%%%%%%%%%%%%%%%%%%%%%%%%%%%%%%%%%

\bibliographystyle{apsrev4-2}

\begin{thebibliography}{300}
\frenchspacing

\bibitem{peebles} P. J. E. Peebles and B. Ratra, \textit{The Cosmological Constant and Dark Energy}, Rev. Mod. Phys. \textbf{75}, 559 (2003) [arXiv:astro-ph/0207347]

\bibitem{copeland} E. J. Copeland, M. Sami and S. Tsujikawa, \textit{Dynamics of dark energy}, Int. J. Mod. Phys. D \textbf{15}, 1753 (2006) [arXiv:hep-th/0603057].

\bibitem{riess98}
    A.~G.~Riess \textit{et al.} [Supernova Search Team],
    \textit{Observational evidence from supernovae for an accelerating universe and a cosmological constant},
    Astron. J. \textbf{116}, 1009 (1998) [arXiv:astro-ph/9805201].

\bibitem{perlmutter99}
    S.~Perlmutter \textit{et al.} [Supernova Cosmology Project],
    \textit{Measurements of $\Omega$ and $\Lambda$ from 42 high redshift supernovae},
    Astrophys. J. \textbf{517}, 565 (1999) [arXiv:astro-ph/9812133].

\bibitem{planck21} 
  N.~Aghanim {\it et al.} [Planck Collaboration],
   \textit{Planck 2018 results. VI. Cosmological parameters},
   Astron. Astrophys. \textbf{641}, A6  (2020) [erratum: Astron. Astrophys. \textbf{652}, C4 (2021)] [arXiv:1807.06209].

\bibitem{pantheon18} 
  D.~M.~Scolnic {\it et al.},
  \textit{The Complete Light-curve Sample of Spectroscopically Confirmed SNe Ia from Pan-STARRS1 and Cosmological Constraints from the Combined Pantheon Sample},
  Astrophys.\ J.\  {\bf 859}, 101 (2018)
  [arXiv:1710.00845].

\bibitem{eboss21}
    S.~Alam \textit{et al.} [eBOSS],
    \textit{Completed SDSS-IV extended Baryon Oscillation Spectroscopic Survey: Cosmological implications from two decades of spectroscopic surveys at the Apache Point Observatory},
    Phys. Rev. D \textbf{103}, 083533 (2021)
    [arXiv:2007.08991].

\bibitem{kids21}
    C.~Heymans \textit{et al.}
    \textit{KiDS-1000 Cosmology: Multi-probe weak gravitational lensing and spectroscopic galaxy clustering constraints},
    Astron. Astrophys. \textbf{646}, A140 (2021) 
    [arXiv:2007.15632].
    
\bibitem{des21a}
    T.~M.~C.~Abbott \textit{et al.} [DES],
    \textit{Dark Energy Survey Year 3 Results: Cosmological Constraints from Galaxy Clustering and Weak Lensing},
    Phys. Rev. D \textbf{105}, 023520 (2022) 
    [arXiv:2105.13549].

\bibitem{des21b}
    L.~F.~Secco \textit{et al.} [DES],
    \textit{Dark Energy Survey Year 3 Results: Cosmology from Cosmic Shear and Robustness to Modeling Uncertainty},
    Phys. Rev. D \textbf{105}, 023515  (2022) 
    [arXiv:2105.13544].

\bibitem{weinberg89}
    S.~Weinberg,
    \textit{The Cosmological Constant Problem}, Rev. Mod. Phys. \textbf{61}, 1 (1989).
    
\bibitem{padmanabhan03}
    T.~Padmanabhan,
    \textit{Cosmological constant: The Weight of the vacuum}, Phys. Rept. \textbf{380}, 235 (2003) 
    [arXiv:hep-th/0212290].

\bibitem{baumann18}
    D.~Baumann,
    \textit{Primordial Cosmology},
    PoS \textbf{TASI2017}, 009 (2018) 
    [arXiv:1807.03098].

\bibitem{H0tension} E. Di Valentino \textit{et al.}, 
\textit{Cosmology intertwined II: The Hubble constant tension}, 
Astropart. Phys. \textbf{131}, 102605 (2021) 
[arXiv:2008.11284].

\bibitem{s8tension} E. Di Valentino \textit{et al.}, 
\textit{Cosmology Intertwined III: $f \sigma_8$ and $S_8$}, 
Astropart. Phys. \textbf{131}, 102604 (2021) 
[arXiv:2008.11285].
  
\bibitem{divalentino21}
    E.~Di Valentino, O.~Mena, S.~Pan, L.~Visinelli, W.~Yang, A.~Melchiorri, D.~F.~Mota, A.~G.~Riess and J.~Silk,
    {\it In the realm of the Hubble tension\textemdash{}a review of solutions},
    Class. Quant. Grav. \textbf{38}, 153001  (2021) 
    [arXiv:2103.01183].

\bibitem{shah21}
    P.~Shah, P.~Lemos and O.~Lahav,
    \textit{A buyer's guide to the Hubble Constant},
    Astron. Astrophys. Rev. \textbf{29}, 9  (2021) 
    [arXiv:2109.01161].
    
\bibitem{riess22}
    A.~G.~Riess \textit{et al.}
    \textit{A Comprehensive Measurement of the Local Value of the Hubble Constant with 1 km/s/Mpc Uncertainty from the Hubble Space Telescope and the SH0ES Team},
    Astrophys. J. Lett. \textbf{934}, L7 (2022) 
    [arXiv:2112.04510].


\bibitem{uzan} J.-P. Uzan, \textit{Varying Constants, Gravitation and Cosmology}, Living Rev. Relativ. \textbf{14}, 2 (2011) [arXiv:1009.5514].


\bibitem{dicke57}
    Dicke, R.~H.
     \textit{Gravitation without a Principle of Equivalence},
     Reviews of Modern Physics (1957)

\bibitem{moffat93}
    J.~W.~Moffat,
    \textit{Superluminary universe: A Possible solution to the initial value problem in cosmology}, 
    Int. J. Mod. Phys. D \textbf{2}, 351 (1993) 
    [arXiv:gr-qc/9211020].

\bibitem{avelino99}
    P.~P.~Avelino and C.~J.~A.~P.~Martins,
    \textit{Does a varying speed of light solve the cosmological problems?},
    Phys. Lett. B \textbf{459}, 468 (1999) 
    [arXiv:astro-ph/9906117].

\bibitem{magueijo00}
    J.~Magueijo,
    \textit{Covariant and locally Lorentz invariant varying speed of light theories},
    Phys. Rev. D \textbf{62}, 103521 (2000) 
    [arXiv:gr-qc/0007036].

\bibitem{moffat16}
    J.~W.~Moffat,
    \textit{Variable Speed of Light Cosmology, Primordial Fluctuations and Gravitational Waves},
    Eur. Phys. J. C \textbf{76}, 130  (2016) 
    [arXiv:1404.5567].

\bibitem{barrow98}
    J.~D.~Barrow,
    \textit{Cosmologies with varying light speed},
    [arXiv:astro-ph/9811022].

\bibitem{albrecht99}
    A.~Albrecht and J.~Magueijo,
    \textit{A Time varying speed of light as a solution to cosmological puzzles},
    Phys. Rev. D \textbf{59}, 043516  (1999) 
    [arXiv:astro-ph/9811018].

\bibitem{barrow99a}
    J.~D.~Barrow and J.~Magueijo,
    \textit{Solutions to the quasi-flatness and quasilambda problems},
    Phys. Lett. B \textbf{447}, 246 (1999) 
    [arXiv:astro-ph/9811073].

\bibitem{clayton99}
    M.~A.~Clayton and J.~W.~Moffat,
    \textit{Dynamical mechanism for varying light velocity as a solution to cosmological problems},
    Phys. Lett. B \textbf{460}, 263 (1999) 
    [arXiv:astro-ph/9812481].

\bibitem{barrow99b}
    J.~D.~Barrow and J.~Magueijo,
    \textit{Solving the flatness and quasiflatness problems in Brans-Dicke cosmologies with a varying light speed},
    Class. Quant. Grav. \textbf{16}, 1435 (1999) 
    [arXiv:astro-ph/9901049].

\bibitem{clayton00}
    M.~A.~Clayton and J.~W.~Moffat,
    \textit{Scalar tensor gravity theory for dynamical light velocity},
    Phys. Lett. B \textbf{477}, 269 (2000) 
%    doi:10.1016/S0370-2693(00)00192-1
    [arXiv:gr-qc/9910112].

\bibitem{bassett00}
    B.~A.~Bassett, S.~Liberati, C.~Molina-Paris and M.~Visser,
    \textit{Geometrodynamics of variable speed of light cosmologies},
    Phys. Rev. D \textbf{62}, 103518  (2000) 
    [arXiv:astro-ph/0001441].

\bibitem{clayton02}
    M.~A.~Clayton and J.~W.~Moffat,
    \textit{Vector field mediated models of dynamical light velocity},
    Int. J. Mod. Phys. D \textbf{11}, 187  (2002) 
    [arXiv:gr-qc/0003070].

\bibitem{magueijo03}
    J.~Magueijo,
    \textit{New varying speed of light theories}, 
    Rept. Prog. Phys. \textbf{66}, 2025 (2003) 
    [arXiv:astro-ph/0305457].

\bibitem{ellis05}
    G.~F.~R.~Ellis and J.~P.~Uzan,
    \textit{`c' is the speed of light, isn't it?},
    Am. J. Phys. \textbf{73}, 240-247 (2005) 
%    doi:10.1119/1.1819929
    [arXiv:gr-qc/0305099].
    
\bibitem{ellis07}
    G.~F.~R.~Ellis,
    \textit{Note on Varying Speed of Light Cosmologies},
    Gen. Rel. Grav. \textbf{39}, 511  (2007) 
    [arXiv:astro-ph/0703751].

\bibitem{magueijo08}
    J.~Magueijo and J.~W.~Moffat,
    \textit{Comments on `Note on varying speed of light theories'},
    Gen. Rel. Grav. \textbf{40}, 1797 (2008) 
    [arXiv:0705.4507].
    
\bibitem{cruz12}
    C.~N.~Cruz and A.~C.~A.~d.~Faria,
    \textit{Variation of the speed of light with temperature of the expanding universe},
    Phys. Rev. D \textbf{86}, 027703  (2012) 
    [arXiv:1205.2298].

\bibitem{franzmann17}
    G.~Franzmann,
    \textit{Varying fundamental constants: a full covariant approach and cosmological applications},
    [arXiv:1704.07368].

\bibitem{cruz18}
    C.~N.~Cruz and F.~A.~da Silva,
    \textit{Variation of the speed of light and a minimum speed in the scenario of an inflationary universe with accelerated expansion},
    Phys. Dark Univ. \textbf{22}, 127 (2018) 
    [arXiv:2009.05397].

\bibitem{costa19}
    R.~Costa, R.~R.~Cuzinatto, E.~M.~G.~Ferreira and G.~Franzmann,
    \textit{Covariant c-flation: a variational approach},
    Int. J. Mod. Phys. D \textbf{28}, 1950119  (2019) 
    [arXiv:1705.03461].

\bibitem{gupta20}
    R.~P.~Gupta,
    \textit{Cosmology with relativistically varying physical constants},
    Mon. Not. Roy. Astron. Soc. \textbf{498}, 4481  (2020) 
    [arXiv:2009.08878].

\bibitem{lee21a}
    S.~Lee,
    \textit{The minimally extended Varying Speed of Light (meVSL)},
    JCAP \textbf{08}, 054  (2021) 
    [arXiv:2011.09274].
    
\bibitem{lee21b}
    S.~Lee,
    \textit{Constraints on the time variation of the speed of light using Pantheon dataset},
    [arXiv:2101.09862].

\bibitem{lee21c}
    S.~Lee,
    \textit{Constraints on the time variation of the speed of light using Strong lensing},
    [arXiv:2104.09690].

\bibitem{lee21d}
    S.~Lee,
    \textit{Cosmic distance duality as a probe of minimally extended varying speed of light},
    [arXiv:2108.06043].

\bibitem{lee21e}
    S.~Lee,
    \textit{Determination of varying speed of light from Black hole},
    [arXiv:2110.08809].

\bibitem{cuzinatto22}
    R.~R.~Cuzinatto, R.~P.~Gupta, R.~F.~L.~Holanda, J.~F.~Jesus and S.~H.~Pereira,
    \textit{Testing a varying-\ensuremath{\Lambda} model for dark energy within co-varying physical couplings framework},
    Mon. Not. Roy. Astron. Soc. \textbf{515}, 5981 (2022)
    [arXiv:2204.10764].

\bibitem{lee23}
    S.~Lee,
    \textit{Constraining minimally extended varying speed of light by cosmological chronometers},
    [arXiv:2301.06947].

\bibitem{gupta23}
    R.~P.~Gupta,
    \textit{Constraining Coupling Constants' Variation with Supernovae, Quasars, and GRBs},
    Symmetry \textbf{15} (2023), 259
    [arXiv:2301.09795].

\bibitem{salzano15}
    V.~Salzano, M.~P.~Dabrowski and R.~Lazkoz,
    \textit{Measuring the speed of light with Baryon Acoustic Oscillations},
    Phys. Rev. Lett. \textbf{114}, 101304  (2015) 
    [arXiv:1412.5653].

\bibitem{cao17}
    S.~Cao, M.~Biesiada, J.~Jackson, X.~Zheng, Y.~Zhao and Z.~H.~Zhu,
    \textit{Measuring the speed of light with ultra-compact radio quasars},
    JCAP \textbf{02}, 012  (2017) 
    [arXiv:1609.08748].

\bibitem{gabriel22}
    G.~Rodrigues and C.Bengaly,
     \textit{A model-independent test of speed of light variability with cosmological observations},
    JCAP \textbf{7}, 029  (2022) 
    [arXiv:2112.01963].

\bibitem{salzano17}
     V.~Salzano,
    \textit{Recovering a redshift-extended varying speed of light signal from galaxy surveys},
    Phys. Rev. D \textbf{95}, 084035  (2017) 
    [arXiv:1604.03398].

\bibitem{cai16}
    R.~G.~Cai, Z.~K.~Guo and T.~Yang,
    \textit{Dodging the cosmic curvature to probe the constancy of the speed of light},
    JCAP \textbf{08}, 016  (2016) 
    [arXiv:1601.05497].

\bibitem{cc0} R. Jimenez and A. Loeb, \textit{Constraining cosmological parameters based on relative galaxy ages},
Astrophys. J. \textbf{573}, 37  (2002) [arXiv:astro-ph/0106145].

\bibitem{cc1} R. Jimenez, L. Verde, T. Treu and D. Stern, \textit{Constraints on the equation of state of dark
energy and the Hubble constant from stellar ages and the CMB}, Astrophys. J. \textbf{593}, 622 (2003) 
[arXiv:astro-ph/0302560].

\bibitem{cc2} J. Simon, L. Verde and R. Jimenez, \textit{Constraints on the redshift dependence of the dark energy
potential}, Phys. Rev. D \textbf{71}, 123001 (2005) [arXiv:astro-ph/0412269].

\bibitem{cc3} D. Stern, R. Jimenez, L. Verde, M. Kamionkowski and S.A. Stanford, \textit{Cosmic Chronometers:
Constraining the Equation of State of Dark Energy. I: $H(z)$ Measurements}, JCAP \textbf{02}, 008 (2010) [arXiv:0907.3149].


\bibitem{cc4} M. Moresco \textit{et al.}, \textit{Improved constraints on the expansion rate of the Universe up to $z \sim 1.1$
from the spectroscopic evolution of cosmic chronometers}, JCAP \textbf{08}, 006  (2012) [arXiv:1201.3609].

\bibitem{cc5} C. Zhang, H. Zhang, S. Yuan, T.-J. Zhang and Y.-C. Sun, \textit{Four new observational $H(z)$ data 
from luminous red galaxies in the Sloan Digital Sky Survey data release seven}, Res. Astron.
Astrophys. \textbf{14}, 1221 (2014) [arXiv:1207.4541].

\bibitem{cc6} M. Moresco, \textit{Raising the bar: new constraints on the Hubble parameter with cosmic
chronometers at $z \sim 2$}, Mon. Not. Roy. Astron. Soc. \textbf{450}, L16 (2015) [arXiv:1503.01116].

\bibitem{cc7} M. Moresco, L. Pozzetti, A. Cimatti, R. Jimenez, C. Maraston, L. Verde \textit{et al.}, \textit{A 6\%
measurement of the Hubble parameter at $z \sim 0.45$: direct evidence of the epoch of cosmic
re-acceleration}, JCAP \textbf{05}, 014  (2016) [arXiv:1601.01701].

\bibitem{cc8} A. L. Ratsimbazafy, S. I. Loubser, S. M. Crawford, C. M. Cress, B. A. Bassett, R. C. Nichol \textit{et al.},
\textit{Age-dating Luminous Red Galaxies observed with the Southern African Large Telescope}, Mon. Not. Roy. Astron. Soc. \textbf{467}, 3239 (2017) [arXiv:1702.00418].

\bibitem{cc9} N. Borghi, M. Moresco and A. Cimatti, \textit{Toward a Better Understanding of Cosmic
Chronometers: A New Measurement of $H(z)$ at $z \sim 0.7$}, Astrophys. J. Lett. \textbf{928}, L4  (2022) 
[arXiv:2110.04304].

\bibitem{bao1} G.-B. Zhao \textit{et al.}, \textit{The clustering of the SDSS-IV extended Baryon Oscillation Spectroscopic
Survey DR14 quasar sample: a tomographic measurement of cosmic structure growth and
expansion rate based on optimal redshift weights}, Mon. Not. Roy. Astron. Soc. \textbf{482}, 3497 (2019) [arXiv:1801.03043].

\bibitem{bao2} E. Gaztanaga, A. Cabre and L. Hui, \textit{Clustering of Luminous Red Galaxies IV: Baryon
Acoustic Peak in the Line-of-Sight Direction and a Direct Measurement of $H(z)$}, Mon. Not.
Roy. Astron. Soc. \textbf{399}, 1663  (2009) [arXiv:0807.3551].

\bibitem{bao3} C. Blake \textit{et al.}, \textit{The WiggleZ Dark Energy Survey: Joint measurements of the expansion and
growth history at $z < 1$}, Mon. Not. Roy. Astron. Soc. \textbf{425}, 405 (2012) [arXiv:1204.3674].

\bibitem{bao4} L. Samushia \textit{et al.}, \textit{The Clustering of Galaxies in the SDSS-III DR9 Baryon Oscillation
Spectroscopic Survey: Testing Deviations from $\Lambda$ and General Relativity using anisotropic
clustering of galaxies}, Mon. Not. Roy. Astron. Soc. \textbf{429}, 1514 (2013) [arXiv:1206.5309].

\bibitem{bao5} X. Xu, A.J. Cuesta, N. Padmanabhan, D.J. Eisenstein and C.K. McBride, \textit{Measuring $D_A$ and
$H$ at $z = 0.35$ from the SDSS DR7 LRGs using baryon acoustic oscillations}, Mon. Not. Roy.
Astron. Soc. \textbf{431}, 2834 (2013) [arXiv:1206.6732].

\bibitem{bao6} T. Delubac \textit{et al.} [BOSS collaboration], \textit{Baryon acoustic oscillations in the Ly$\alpha$ forest of BOSS DR11 quasars},
Astron. Astrophys. \textbf{574}, A59 (2015) [arXiv:1404.1801].

\bibitem{bao7} A. Font-Ribera \textit{et al.} [BOSS collaboration], \textit{Quasar-Lyman $\alpha$ Forest Cross-Correlation from BOSS DR11 : Baryon
Acoustic Oscillations}, JCAP \textbf{05}, 027 (2014) [arXiv:1311.1767].

\bibitem{bao8} S. Alam \textit{et al.} [BOSS collaboration], \textit{The clustering of galaxies in the completed SDSS-III Baryon Oscillation
Spectroscopic Survey: cosmological analysis of the DR12 galaxy sample}, Mon. Not. Roy.
Astron. Soc. \textbf{470}, 2617 (2017) [arXiv:1607.03155].

\bibitem{bao9} H. du Mas des Bourboux \textit{et al.}, \textit{Baryon acoustic oscillations from the complete SDSS-III
Ly$\alpha$-quasar cross-correlation function at $z = 2.4$}, Astron. Astrophys. \textbf{608}, A130 (2017) 
[arXiv:1708.02225].

\bibitem{nunes20}
    R.~C.~Nunes, S.~K.~Yadav, J.~F.~Jesus and A.~Bernui,
    \textit{Cosmological parameter analyses using transversal BAO data},
    Mon. Not. Roy. Astron. Soc. \textbf{497}, 2133 (2020) 
    [arXiv:2002.09293].
    
\bibitem{carvalho16}
    G.~C.~Carvalho, A.~Bernui, M.~Benetti, J.~C.~Carvalho and J.~S.~Alcaniz,
    \textit{Baryon Acoustic Oscillations from the SDSS DR10 galaxies angular correlation function},
    Phys. Rev. D \textbf{93}, 023530  (2016) 
    [arXiv:1507.08972].

\bibitem{alcaniz17}
    J.~S.~Alcaniz, G.~C.~Carvalho, A.~Bernui, J.~C.~Carvalho and M.~Benetti,
    \textit{Measuring baryon acoustic oscillations with angular two-point correlation function},
    Fundam. Theor. Phys. \textbf{187}, 11  (2017) 
    [arXiv:1611.08458].

\bibitem{carvalho20}
    G.~C.~Carvalho, A.~Bernui, M.~Benetti, J.~C.~Carvalho, E.~de Carvalho and J.~S.~Alcaniz,
    \textit{The transverse baryonic acoustic scale from the SDSS DR11 galaxies}, 
    Astropart. Phys. \textbf{119}, 102432  (2020) 
    [arXiv:1709.00271].

\bibitem{decarvalho18}
    E.~de Carvalho, A.~Bernui, G.~C.~Carvalho, C.~P.~Novaes and H.~S.~Xavier,
    \textit{Angular Baryon Acoustic Oscillation measure at $z=2.225$ from the SDSS quasar survey},
    JCAP \textbf{04}, 064 (2018) 
    [arXiv:1709.00113].

\bibitem{decarvalho21}
    E.~de Carvalho, A.~Bernui, F.~Avila, C.~P.~Novaes and J.~P.~Nogueira-Cavalcante,
    \textit{BAO angular scale at $z_{\text{eff}} = 0.11$ with the SDSS blue galaxies},
    Astron. Astrophys. \textbf{649}, A20  (2021) 
    [arXiv:2103.14121].

\bibitem{kazantzidis20}
    L.~Kazantzidis and L.~Perivolaropoulos,
    ``Hints of a Local Matter Underdensity or Modified Gravity in the Low $z$ Pantheon data,''
    Phys. Rev. D \textbf{102}, no.2, 023520 (2020)
    [arXiv:2004.02155].

\bibitem{ruchika23}
    Ruchika, H.~Rathore, S.~Roy Choudhury and V.~Rentala,
    ``A gravitational constant transition within cepheids as supernovae calibrators can solve the Hubble tension,''
    [arXiv:2306.05450].

%\bibitem{cc_sys}
%	M. Moresco, R. Jimenez, L. Verde, A. Cimatti and L. Pozzetti, 
%	{\it Setting the Stage for Cosmic Chronometers. II. Impact of Stellar Population Synthesis Models Systematics and Full Covariance Matrix}, 
%	Astrophys. J. \textbf{898}, 82 (2020) 
%	[arXiv:2003.07362].

\bibitem{seikel12} 
  M.~Seikel, C.~Clarkson and M.~Smith,
   \textit{Reconstruction of dark energy and expansion dynamics using Gaussian processes}, 
  JCAP {\bf 06}, 036 (2012)
  [arXiv:1204.2832].\\
  GaPP is available at \url{https://github.com/astrobengaly/GaPP}
  
\bibitem{shafieloo12} 
    A.~Shafieloo, A.~G.~Kim and E.~V.~Linder,
    \textit{Gaussian Process Cosmography},
    Phys.\ Rev.\ D {\bf 85}, 123530 (2012)
    [arXiv:1204.2272].

\bibitem{eoin} E. \'{O} Colg\'{a}in and M. M. Sheikh-Jabbari, \textit{Elucidating cosmological model dependence with $H_0$}, Eur. Phys. J. C \textbf{81}, 892 (2021) [arXiv:2101.08565].

\bibitem{purba_j} P. Mukherjee and N. Banerjee, \textit{Non-parametric reconstruction of the cosmological \textit{jerk} parameter}, Eur. Phys. J. C \textbf{81}, 36 (2021) [arXiv:2007.10124].

\bibitem{purba_q} P. Mukherjee and N. Banerjee, \textit{Revisiting a non-parametric reconstruction of the deceleration parameter from combined background and the growth rate data}, Phys. Dark Univ. \textbf{36}, 100998 (2022) [arXiv:2007.15941].


\bibitem{riess21} A. G. Riess, S. Casertano, W. Yuan, J.B. Bowers, L. Macri, J.C. Zinn \textit{et al.}, \textit{Cosmic Distances Calibrated to 1\% Precision with Gaia EDR3 Parallaxes and Hubble Space Telescope Photometry of 75 Milky Way Cepheids Confirm Tension with $\Lambda$CDM}, Astrophys. J. Lett. \textbf{908}, L6  (2021) [arXiv:2012.08534].

\bibitem{freedman21}
    W.~L.~Freedman,
    \textit{Measurements of the Hubble Constant: Tensions in Perspective},
    Astrophys. J. \textbf{919}, 16  (2021) 
    [arXiv:2106.15656].

\bibitem{efstathiou21}
    G.~Efstathiou,
    \textit{To H0 or not to H0?},
    Mon. Not. Roy. Astron. Soc. \textbf{505}, 3866 (2021) 
    [arXiv:2103.08723].

\bibitem{camarena21}
    D.~Camarena and V.~Marra,
    \textit{On the use of the local prior on the absolute magnitude of Type Ia supernovae in cosmological inference}, 
    Mon. Not. Roy. Astron. Soc. \textbf{504}, 5164 (2021) 
    [arXiv:2101.08641].

\bibitem{camarena20}
    D.~Camarena and V.~Marra,
    \textit{A new method to build the (inverse) distance ladder}, 
    Mon. Not. Roy. Astron. Soc. \textbf{495}, 2630  (2020) 
    [arXiv:1910.14125].

\bibitem{pavon} L.~P. Chimento, A.~S. Jakubi, and D. Pav\'{o}n,
{\it Varying $c$ and particle horizons},
Phys. Lett. B,
\textbf{508}, 1 (2001)
[arXiv:gr-qc/0103038].


\bibitem{jpas14}
    N.~Benitez \textit{et al.} [J-PAS],
    \textit{J-PAS: The Javalambre-Physics of the Accelerated Universe Astrophysical Survey},
    [arXiv:1403.5237].

\bibitem{euclid18} 
  L.~Amendola {\it et al.},
   \textit{Cosmology and fundamental physics with the Euclid satellite},
  Living Rev.\ Rel.\  {\bf 21},  2 (2018)
   [arXiv:1606.00180].
  
\bibitem{ska20} 
  D.~J.~Bacon {\it et al.} [SKA Collaboration],
   \textit{Cosmology with Phase 1 of the Square Kilometre Array: Red Book 2018: Technical specifications and performance forecasts},
  Publ. Astron. Soc. Austral. \textbf{37} (2020), e007
  [arXiv:1811.02743].
  
\bibitem{lsst18}
  D.~Alonso \textit{et al.} [LSST Dark Energy Science],
  \textit{The LSST Dark Energy Science Collaboration (DESC) Science Requirements Document},
  [arXiv:1809.01669].

\bibitem{punturo10}
    M.~Punturo \textit{et al.}
    \textit{The third generation of gravitational wave observatories and their science reach},
    Class. Quant. Grav. \textbf{27} (2010), 084007



\end{thebibliography}

%%%%%%%%%%%%%%%%%%%%%%%%%%%%%%%%%%%%%%%%%%%%%%%%%%%

\label{lastpage}

\end{document}